\documentclass[a4paper,11pt]{article}
\usepackage{jcappub}
\usepackage[T1]{fontenc}
\usepackage[utf8]{inputenc}
\usepackage{braket}
\usepackage{bm}
\usepackage{tikz}
\usepackage{mathtools}
\usepackage{xcolor}
\usepackage{soul}
\usepackage{comment}
\usepackage{subcaption} 
\usepackage{booktabs}
\usepackage{float}
\usepackage{amsmath,amssymb}
\usepackage{verbatim}

\newcommand\numberthis{\addtocounter{equation}{1}\tag{\theequation}}

\def\f12{\frac{1}{2}}
\def\f13{\frac{1}{3}}

\newcommand{\aB}{\alpha_{\rm B}}
\newcommand{\aK}{\alpha_{\rm K}}
\newcommand{\aT}{\alpha_{\rm T}}
\newcommand{\aH}{\alpha_{\rm H}}
\newcommand{\aM}{\alpha_{\rm M}}
\everymath{\displaystyle}

\title{Linear perturbation theory and structure formation in a Horndeski theory of gravity without dark matter}

\author[a,b]{Lorenzo Gervani,}
\author[a,b,c]{Francesco Pace,}
\author[a,b,d]{Antonaldo Diaferio}

\affiliation[a]{Dipartimento di Fisica, Universit\`a degli Studi di Torino, Via P. Giuria 1, I-10125, Torino, Italy}
\affiliation[b]{INFN - Sezione di Torino, Via P. Giuria 1, I-10125, Torino, Italy}
\affiliation[c]{INAF - Istituto Nazionale di Astrofisica, Osservatorio Astrofisico di Torino, strada Osservatorio 20, 10025, Pino Torinese, Italy}
\affiliation[d]{Accademia delle Scienze di Torino, Via Maria Vittoria 3, I-10123, Torino, Italy}

\emailAdd{lorenzo.gervani@unito.it}
\emailAdd{antonaldo.diaferio@unito.it}
\emailAdd{francesco.pace@unito.it}

\abstract{
    We study the formation and evolution of the cosmic large-scale structure (LSS) in a scalar-tensor Horndeski theory of gravity without dark matter. We impose that this theory reduces to Refracted Gravity (RG) in the weak-field limit. RG is a non-relativistic theory where the gravitational permittivity $\epsilon(\rho)$, with $\rho$ the baryonic matter density, provides the gravitational boost required to describe both the dynamics of galaxies and galaxy clusters without dark matter. We show that $\epsilon(\rho)$ equals the function $G_4(\varphi)$ of the scalar field $\varphi$  of our Horndeski model.  
    The background evolution of a flat, matter-dominated, homogeneous and isotropic universe shows that $\varphi$ is also responsible for the accelerated expansion of the universe and that the model is capable of reproducing the observed redshift evolution of the Hubble-Lema\^\i tre parameter. 
    We finally develop the linear perturbation theory and study the growth factor equation. 
    The model fails to correctly describe the LSS by predicting a late onset of  the formation of cosmic structure  at redshift $z\lesssim 1$. 
}

\begin{document}

\maketitle
\flushbottom
\allowdisplaybreaks

\section{Introduction}
To date, General Relativity (GR) remains the best available theory  to describe gravitational phenomena. Its many successes include, at cosmological scales, those achieved by the $\Lambda$ cold dark matter ($\Lambda$CDM) model, where an unspecified form of collisionless non-baryonic cold dark matter (CDM) is introduced. In addition, alongside dark matter (DM), a cosmological constant $\Lambda$ is required to explain the observed present-day acceleration in the expansion rate of the universe. CDM plays a crucial role in the model: it is invoked to accommodate a plethora of different observations, spanning from the smallest scales of galaxies 
— the so-called "missing mass" or "missing gravity" problem \cite{deMartino:2020gfi, Iocco:2015xga} — to the largest cosmic scales — where the list of explained phenomena includes, but is not limited to, the formation of the large-scale structure \cite{Clowe:2006eq, SDSS:2001afu, DelPopolo:2013qba}, the abundances of light elements \cite{Cyburt:2015mya, Fields:2019pfx}, the anisotropies of temperature fluctuations in the cosmic microwave background (CMB) \cite{Planck:2018vyg, Planck:2019nip} and the measured amplitude of the weak gravitational lensing effect \cite{Bartelmann:1999yn}. Without CDM, all of these observations would be difficult to explain in the context of a universe governed by GR. With CDM, the $\Lambda$CDM model is able to give a posteriori estimates of the six main cosmological parameters 
that describe 
several observables, including the CMB anisotropies spectrum \cite{Planck:2019nip, Wallisch:2018rzj} and the evolution of the expansion rate \cite{Riess_2024, Freedman_2025}. 
The measurement of dynamical observables enables to infer the amount of DM that must be present around a specific object, like a galaxy or a galaxy cluster, in order to solve the missing mass problem \cite{Navarro:1996gj, Salucci:2018eie, Salucci:2018hqu}. Numerical simulations in the context of the $\Lambda$CDM model track the formation and evolution of the large-scale structure that appear to agree with the real universe \cite{Kuhlen:2012ft}. 

Despite this alleged success, over the years, $\Lambda$CDM has accumulated a certain number of tensions, both on galactic and cosmic scales. One of the most investigated tensions concerns the Hubble-Lemaître parameter $H_0$: the value estimated with the distance ladder method \cite{Riess_2024, H0DN:2025lyy} significantly differs from the best-fit value derived from the analysis of the CMB spectrum \cite{Freedman_2025}. Another important cosmological tension concerns the observed and predicted abundance of $^7$Li \cite{Mathews:2019hbi, Fields:2022mpw, Planck:2019nip}. In addition, there exists a strong tension between the observed value of the cosmological constant $\Lambda$ and the value expected from the usual quantum field theory interpretation of $\Lambda$ as being associated to the vacuum energy level \cite{Weinberg:1988cp, Padilla:2015aaa}. Recent measures seem to suggest that $\Lambda$ might not be a constant at all \cite{DESI:2025zgx}. On smaller scales, the numerical results obtained from N-body simulations of structure formation in $\Lambda$CDM also face a number of challenges, including the core-cusp problem~\cite{McGaugh:1998tq}, the dwarf galaxy problem~\cite{Moore:1999nt, Klypin:1999uc}, the too-big-to-fail problem~\cite{Boylan-Kolchin:2011qkt} and the plane of satellite problem~\cite{Pawlowski:2014una}. 

An additional relevant weakness of the model concerns several phenomenological relations observed in galactic dynamics. The model does not naturally  account for the radial acceleration relation (RAR) — which relates the measured acceleration of stars in disk galaxies and the acceleration due to the \textit{baryonic} mass component of the galaxy \cite{Lelli:2016cui} — and for the baryonic Tully-Fisher relation (BTFR) — the tight power-law correlation between a galaxy's total baryonic mass and its rotation velocity \cite{Tully:1977fu, McGaugh:2000sr}. The DM model can explain some of these problematic aspects, but only at the cost of a careful fine-tuning of the parameters governing  galaxy formation \cite{deMartino:2020gfi}: for instance, simulations can reproduce the BTFR by assuming very precise relations between the DM and the baryonic content of the halo \cite{DiCintio:2015eeq, Lelli_2016}. Still, these relations possess a statistical scatter that is larger than the very tight observed scatter  \cite{McGaugh_2012, scannapieco2012aquila, Ponomareva_2017,Lelli:2016cui}. These discrepancies may be explained by assuming a CDM halo distribution different from the classical Navarro-Frenk-White profile \cite{Gentile:2004tb}, or by invoking complicated "gastrophysical" feedback effects \cite{galaxies5010017, Papastergis_2016}.

A final challenge to the DM paradigm is the fact that, despite decades of efforts, no clear indication of the existence of a DM particle has been obtained yet. Weakly interactive massive particles (WIMPs) have long been considered the most likely candidate for DM, and more recently many other types of candidate have been studied, including axions, sterile neutrinos, primordial black holes, and hidden sector DM \cite{deMartino:2020gfi, Bullock:2017xww, Famaey:2011kh}. All of these possibilities are currently under scrutiny, with improving bounds on their mass and cross-section, but all still await a confirmation \cite{cirelli2025darkmatter, Misiaszek_2024, Cebri_n_2023}. 

A number of experimental facts, most famously the rotational velocity of stars in the outskirts of disk galaxies, demonstrate that the Newtonian theory of gravitation is unable to correctly predict the internal dynamics of galaxies. The theory of Newton seems to predict "less gravity" than what is needed to explain observations, thus we speak of a "missing gravity" problem \cite{McGaugh:2016leg}. DM is usually invoked to explain these discrepancies. The idea of Modified Gravity (MG) constitutes an alternative way to deal with the inadequacies of the Newtonian prediction: these inadequacies could derive from our fundamental lack of understanding of the theory of gravity, rather than from the failure to account for a new, exotic kind of matter. MOdified Newtonian Dynamics (MOND) \cite{Milgrom:1983ca} constitutes the archetype of MG theories and has proven to be capable of correctly describing and even predicting the dynamics of objects on galactic scale \cite{Milgrom:1983pn, Milgrom:1983zz, McGaugh_2012, McGaugh:2016leg, Lelli:2016cui}. 
MOND is a research program including an extended family of non-relativistic and relativistic theories \cite{Lelli2026}. The latter are 
essential to study the cosmological predictions of the theory \cite{Famaey:2011kh}. Even as the landscape of relativistic theories expanded, it has been extremely challenging to provide a covariant theory capable of reproducing 
the observational successes of $\Lambda$CDM on cosmic scales \cite{Famaey:2011kh, Skordis:2020eui}. For an extensive review of current observational constraints on various MG models, we refer the reader to \cite{Euclid:2025vml}. 

In our previous work \cite{Gervani:2026zqm}, we considered a MG theory belonging to the class of the Brans--Dicke theories, with the aim of studying whether such a theory was capable of reproducing the expected behavior for the growth factor of density fluctuations in an expanding universe, without the need of introducing a DM component. Brans--Dicke theories are a class of scalar-tensor theories, where a new degree of freedom is introduced in the form of a scalar field, with an associated kinetic term and a scalar potential. In that first paper, by fixing the form of potential and kinetic term, we selected a specific Brans--Dicke theory with the requirement that, in its weak-field limit, the theory reduces to Refracted Gravity (RG). RG is a  non-relativistic theory of gravity that appears to reproduce the observed dynamics of galaxies and galaxy clusters ~\cite{Matsakos, Cesare:2020pyz, Cesare:2022khd, Pizzuti}. Specifically, RG solves the missing mass problem through the introduction of a gravitational permittivity, a monotonically increasing function $\epsilon(\rho)$ of the baryonic matter density $\rho$, which provides a boost to the intensity of the gravitational field in low-density environments. It turns out  that the permittivity of the non-relativistic theory coincides with the scalar field divided by 2 in the covariant formulation  \cite{Sanna}. In addition, an effective approach can show  that the scalar field is responsible for the accelerated expansion of the universe \cite{Sanna}. 

Thus, RG could, in principle, explain both dark energy effects on cosmological scales and DM effects on galactic scales. Our first work \cite{Gervani:2026zqm}, however, showed that this Brans--Dicke covariant formulation of RG fails to correctly reproduce the growth of cosmological structure. The theory predicts a very delayed formation of structure, at redshifts $z<1$, a result that is at odds with the observations of galaxies at redshifts  $z>10$ \cite{Sakstein:2017xjx},  although even $\Lambda$CDM struggles to explain such observations \cite{Boylan-Kolchin_2023, Gupta:2023}. This result might be linked to the specific choice of the scalar potential. Here, we check whether it is possible, with a different choice of the potential — that still makes the theory reduce to RG in the weak-field limit — to obtain a result for structure formation in agreement with the observations, or if the Brans--Dicke theory is in fact unable to correctly account for structure formation with \textit{any} potential. In fact, we aim to study an even wider generalization, namely the whole class of Horndeski theories, of which Brans--Dicke theories constitute a sub-class.

In Section \ref{Horndeski theory and generalized Covariant Refracted Gravity} we show that a specific family of Horndeski theories, which we dub generalized CRG (gCRG), reduces to the classical RG formulation in the weak-field regime, i.e. on galactic scales. The covariant theory previously studied in \cite{Gervani:2026zqm} constitutes a single example within this wider class. In Section \ref{bkg} we study the background evolution of a gCRG universe, while in Section \ref{lin pert th} we develop linear perturbation theory and consider structure formation. Section \ref{QSA numerical solution} explains how we numerically solve the perturbation equations using the quasi-static approximation (QSA). In Section \ref{Horndeski theory in a curved universe} we consider whether curvature might have any effect on structure formation in gCRG. We  conclude in Section \ref{Conclusions}.

\section{Horndeski theory and generalized Covariant Refracted Gravity}
\label{Horndeski theory and generalized Covariant Refracted Gravity}
The Horndeski theory represents the most general scalar-tensor theory that still admits second order equations of motion  \cite{Horndeski:1974wa, Kobayashi:2011nu}. The action of this theory can be divided into a matter part and a gravity part as follows
\begin{equation}
   \mathcal{S} = \mathcal{S}_{\text{m}} +  \mathcal{S}_{\text{grav}} = \mathcal{S}_{\text{m}} + \int d^{4} x \sqrt{-g} \mathcal{L}\left(g_{\mu \nu}\right)\,, \quad \mathcal{L} = \sum_{i = 2}^5 \mathcal{L}_i, 
\end{equation} 
where each of the four terms of the Lagrangian density $\mathcal{L}$ is related to a generic function $G_i\left(\varphi, X\right)$ of the scalar field $\varphi$ and its kinetic term $X$ as follows
\begin{align*}
    \begin{array}{l}
\mathcal{L}_{2} \equiv G_{2}(\varphi, X)\,, \\
\mathcal{L}_{3} \equiv G_{3}(\varphi, X) \square \varphi\,, \\
\mathcal{L}_{4} \equiv G_{4}(\varphi, X) R-2 G_{4X}(\varphi, X)\left[(\square \varphi)^{2}-\left(\nabla_{\mu} \nabla_{\nu} \varphi\right)^{2}\right]\,, \\
\mathcal{L}_{5} \equiv G_{5}(\varphi, X) G_{\mu \nu} \nabla^{\mu}\nabla^\nu \varphi+\cfrac{1}{3} G_{5X}(\varphi, X)\left[(\square \varphi)^{3}-3 \square \varphi\left(\nabla_{\mu} \nabla_{\nu} \varphi\right)^{2}+2\left(\nabla_{\mu} \nabla_{\nu} \varphi\right)^{3}\right]\,. \numberthis
\label{Horndeski}
\end{array}
\end{align*}
Here, $X = g^{\mu\nu}\nabla_\mu\varphi\nabla_\nu\varphi$ is the canonical kinetic term of the scalar field, and a subscript indicates differentiation of a $G$ function with respect to $\varphi$ or $X$. Gravitational waves (GW) constraints impose $G_{4X}=G_{5}=0$ \cite{Sakstein:2017xjx, Creminelli:2017sry, Noeller2019}. When $G_4=1/2$ and all the other $G$ functions are null, the theory reduces to classical GR. If one introduces the scalar field $\varphi$ with a minimal coupling, $G_4(\varphi)\propto\varphi$, and at the same time includes a kinetic and a potential term for the scalar field in the function $G_2$ by fixing
\begin{equation}
    G_2(\varphi,X)=\left( - \frac{W(\varphi)}{\varphi} \nabla^{\alpha} \varphi \nabla_{\alpha} \varphi - 2V(\varphi) \right),
    \label{eq:G2}
\end{equation}
 the theory is a Brans--Dicke theory \cite{Brans:1961sx, Quiros}. In Brans--Dicke theories, $G_3=0$. 
 
 The field equations, obtained by variation of the action with respect to the metric and setting $G_{4X}=G_5=0$, are \cite{Kobayashi:2019}:
\begin{align}
    \begin{split}
        & G_{2X}\nabla_\mu\varphi\nabla_\nu\varphi - \frac{1}{2}G_2g_{\mu\nu} + G_{3X}\square\varphi\nabla_\mu\varphi\nabla_\nu\varphi - \nabla_{(\mu}G_3\nabla_{\nu)}\varphi + \frac{1}{2}g_{\mu\nu} \nabla_{\lambda}G_3\nabla^{\lambda}\varphi\\
        & + G_4G_{\mu\nu} + g_{\mu\nu}(G_{4\varphi}\square\varphi + XG_{4\varphi\varphi}) - G_{4\varphi}\nabla_\mu\nabla_\nu\varphi - G_{4\varphi\varphi}\nabla_\mu\varphi\nabla_\nu\varphi = 4\pi G T_{\mu\nu}\quad ,
    \end{split}
    \label{EE}
\end{align}
where we have left implicit the $(\varphi,X)$-dependence of the $G_2$ and $G_3$ functions and the $\varphi$-dependence of the $G_4$ function. $T_{\mu\nu}$ is the stress-energy tensor associated with the energy content of the universe, obtained by the variation of the matter action $\mathcal{S}_{\text{m}}$. The Klein-Gordon (KG) equation for the scalar field is \cite{Kobayashi:2019}:
\begin{equation}
    \nabla^\mu\left[\left(2G_{2X} - 2G_{3X}\square\varphi - 2G_3\frac{\partial\square\varphi}{\partial X} - 2G_{3\varphi}\right)\nabla_\mu\varphi - G_{3X}\nabla_\mu X\right] = G_{2\varphi} - \nabla_\lambda G_{3\varphi}\nabla^\lambda\varphi + G_{4\varphi} R. 
    \label{KG}
\end{equation}

We now ask under which conditions this rather general theory can describe, in its non-relativistic weak-field limit, the small-scale dynamics of galaxies and galaxy clusters according to Refracted Gravity (RG) \cite{Matsakos, Sanna}. 
The dynamics of RG is ruled by a modified version of the Poisson equation
\begin{equation}
    \nabla \cdot \left(\epsilon \nabla \Psi\right) = 4\pi G \rho \,,
    \label{RG equation}
\end{equation}
where $\Psi$ is the gravitational potential, $G$ is the gravitational constant and  $\epsilon = \epsilon\left(\rho\right)$  is the gravitational permittivity. Therefore, we study what constraints must be imposed on the Horndeski functions $G_2, G_3$ and $G_4$ so that the weak field limit of Eqs. (\ref{EE})-(\ref{KG}) corresponds to Eq. (\ref{RG equation}). 

In the weak field limit, the metric is:
\begin{equation}
    {\rm d}s^2 = -(1+2\Psi){\rm d}t^2 + (1+2\Phi)\delta_{ij}{\rm d}x^i{\rm d}x^j
    \label{weak field metric}
\end{equation}
and we consider a static non relativistic fluid for the right-hand side of the Einstein field Eqs. (\ref{EE}). Thus, we assume $\dot{\varphi} = \dot{\Phi} = \dot{\Psi} = 0$. Further, the only non-zero component of the stress-energy tensor is $T_{00}=\rho$. We compute the Christoffel symbols and Ricci tensor components in the metric in Eq.  (\ref{weak field metric}) up to first order in $\Phi$ and $\Psi$. This procedure yields the linearized Einstein field Eqs. (\ref{EE}) and the linearized KG Eq. (\ref{KG}), which can be combined to obtain an expression that relates the Laplacian of the gravitational potential $\Phi$ and the baryonic density $\rho$, to be confronted with Eq. (\ref{RG equation}). The details of this calculation can be found in Appendix \ref{Weak field limit of a Horndeski theory}. 
The end result is that the conditions required to have the Horndeski weak-field equations reduce to the RG Eq. (\ref{RG equation}) are
\begin{equation}
    \left\{ \begin{aligned}
        & G_3 = 0; \quad G_{2XX} =0; \\
        & G_{2}
        - \frac{G_{4}G_{2\varphi}}{G_{4\varphi}}
        + \frac{2G_{4}G_{2\varphi X}}{G_{4\varphi}}X - 2G_{4\varphi\varphi}X = 0; \\
        & \frac{G_4G_{2X}}{G_{4\varphi}} = G_{4\varphi}.
    \end{aligned} \right.
    \label{conditions}
\end{equation}
Now we need to find suitable functions $G_4$ and $G_2$ that satisfy these conditions. From the last condition in Eqs. (\ref{conditions}) we get $G_{2X} = G_{4\varphi}^2/G_4$, which once integrated, recalling that $G_4$ has no X-dependence due to the GWs constraints, yields 
\begin{equation}
    G_2 = X\frac{G_{4\varphi}^2}{G_4} + V(\varphi)\, ,
\end{equation}
The other relevant derivatives of $G_2$ are 
\begin{equation}
    \begin{aligned}
          G_{2\varphi X} & = -\frac{G_{4\varphi}^3}{G_4^2} + \frac{2G_{4\varphi}G_{4\varphi\varphi}}{G_4}; \\
          G_{2\varphi} & = \left(-\frac{G_{4\varphi}^3}{G_4^2} + \frac{2G_{4\varphi}G_{4\varphi\varphi}}{G_4}\right)X + V_\varphi(\varphi).
    \end{aligned}   
\end{equation}
Substituting these expressions for $G_2$ and its derivatives in the third condition of Eqs. (\ref{conditions}) we find $V_\varphi/V = G_{4\varphi}/G_4$, namely $V=2\Xi G_4$, where $\Xi$ is a non-zero constant. This result implies that the choices
\begin{equation}
    G_2(\varphi, X) = \frac{G_{4\varphi}^2}{G_4}X + 2\Xi G_4; \quad G_4(\varphi) = \epsilon(\varphi),
    \label{choices for RG in Horndeski}
\end{equation}
give a Horndeski theory that reduces to RG in the weak-field limit. In our previous work \cite{Gervani:2026zqm}, we considered a specific case of this general result, where the relation between the permittivity and the scalar field was assumed to be linear: if we fix $G_4 = \epsilon = \varphi/2$, we get $G_2 = X/2\varphi + \Xi\varphi$, which is exactly the Brans-Dicke theory with $W(\varphi) = -1$ and $V(\varphi) = -\Xi\varphi$ that was first proposed in \cite{Sanna} and used in \cite{Gervani:2026zqm}.\footnote{We note that the normalization of the gravity part of the Lagrangian density adopted in \cite{Sanna} and \cite{Gervani:2026zqm} is $(16\pi G)^{-1/2}\equiv M_{\rm {pl}}^2/2$, whereas we set $M_{\rm {pl}}=1$ here (compare Eq. (2.4) in \cite{Gervani:2026zqm} and Eq. (\ref{eq:G2}) here). } That theory was dubbed Covariant Refracted Gravity (CRG). 
On the contrary, the theory defined by Eq. (\ref{choices for RG in Horndeski}) reduces to RG for any choice of the permittivity as a function of the scalar field, not just the linear relation $\epsilon(\varphi)\propto \varphi$. We might refer to this theory as generalized CRG (gCRG).

\section{Background universe in gCRG}
\label{bkg}
Here we study the equations for the background evolution of a universe governed by a Horndeski theory in which the choices in Eqs. (\ref{choices for RG in Horndeski}) have been made. We follow the same derivations of our previous work and generalize them. As was shown in \cite{Gervani:2026zqm}, the background evolution of the scalar field can have a great impact on the evolutions of linear perturbations, therefore it is an important preliminary step in our study of the growth factor equation. With Eq. (\ref{choices for RG in Horndeski}), the $00$ and $rr$ components of Eq. (\ref{EE}) and Eq. (\ref{KG}) become, respectively,
\begin{align*}
     & G_4R_{\mu \nu}-\frac{1}{2}\left(\frac{G_{4\varphi}^2}{G_4}X + 2\Xi G_4 + G_4R - 2G_{4\varphi}\square \varphi - 2XG_{4\varphi\varphi}\right) g_{\mu \nu} \\
    &\quad - G_{4\varphi}\nabla_{\mu} \nabla_{\nu} \varphi + \left(\frac{G_{4\varphi}^2}{G_4} - G_{4\varphi\varphi}\right) \nabla_{\mu} \varphi \nabla_{\nu} \varphi = 4 \pi G T_{\mu \nu}\,, \numberthis
    \label{Modified Einstein 2}
\end{align*}
\begin{equation}
    2\frac{G_{4\varphi}^2}{G_4}\square\varphi + \left(-\frac{G_{4\varphi}^3}{G_4^2} + \frac{G_{4\varphi}G_{4\varphi\varphi}}{G_4}\right)X = 2\Xi G_{4\varphi} + G_{4\varphi}R\,.
    \label{Modified KG 2}
\end{equation}
By substituting Eq.~(\ref{Modified KG 2}) into Eq. (\ref{Modified Einstein 2}), we obtain
\begin{equation}
    G_4 R_{\mu \nu} - G_{4\varphi} \nabla_{\mu} \nabla_{\nu} \varphi + \left(\frac{G_{4\varphi}^2}{G_4} - G_{4\varphi\varphi}\right) \nabla_{\mu} \varphi \nabla_{\nu} \varphi = 4 \pi G T_{\mu \nu}\, .
    \label{Rearrenged Modified Einsten}
\end{equation}
Contracting this result with $g^{\mu\nu}$ and substituting into Eq. (\ref{Modified KG 2}) yields
\begin{equation}
    G_{4\varphi}\square \varphi - 2\Xi G_{4} + G_{4\varphi\varphi}X = 4\pi G T\,,
    \label{Rearrenged KG}
\end{equation}
where $T$ is the trace of the stress-energy tensor. We focus on the simplified case of a flat, matter dominated universe, described by the Friedmann-Lemaître-Robertson-Walker (FLRW) metric:
\begin{equation}
    \mathrm{d} s^{2}=-\mathrm{d} t^{2}+a(t)^{2}\left(\frac{\mathrm{d} r^{2}}{1-k r^{2}}+r^{2} \mathrm{d} \theta^{2}+r^{2} \sin ^{2} \theta \mathrm{d} \phi^{2}\right)\,,
    \label{FRW}
\end{equation}
where $a(t)$ is the scale factor, $k$ is the spatial curvature parameter, $t$ is the cosmic coordinate time and $(r,\theta, \phi)$ are the spatial coordinates in the comoving system. By employing the equations of motions in this metric, we are able to derive an expression for the Hubble-Lemaître parameter as a function of the scale factor, $H(a)$. The background evolution of the scalar field $\varphi$ is obtained in an implicit way, so that we can derive the evolution of the boost factor $G_4[\varphi(a)]$ (for details, we refer to Appendix \ref{Background evolution equations in gCRG}). The final results are the following expressions
\begin{equation}
    H(a) = -\frac{1}{2}\Omega_0 H_0 + \sqrt{\frac{1}{4}\Omega_0^2 + 2\Omega_{\Xi 0}}\frac{H_0}{2}\left[ \left(C + \sqrt{C^2 -1}\right)a^{\pm\sqrt{3}} + \left(C + \sqrt{C^2 -1}\right)^{-1}a^{\mp\sqrt{3}} \right]\,,
    \label{H(a) analitico}
\end{equation}
\begin{equation}
     G_4[\varphi(a)]=\frac{H_{0} G_{4,0}}{H(a) a^{3}}.
     \label{G4 analitico}
\end{equation}
The parameter $\Omega_0 \equiv \Omega_{\mathrm{m}0}/G_{4,0} \equiv 8 \pi G\rho_{\mathrm{m}0}/3 H^{2} G_{4,0}$ accounts for the  matter density contribution at the present time. It is rescaled by a factor $1/G_4$ with respect to its $\Lambda$CDM counterpart. This is expected since, as shown above, the function $G_4$ is related to the gravitational permittivity introduced in RG and therefore $G_4$ correctly represents the boost factor. The contribution of dark matter in the $\Lambda$CDM model to the total matter density parameter is replaced here by a boosted baryonic matter density. The parameter $\Omega_{\Xi 0}\equiv\Xi/3 H_0^{2}$, is the analogue of the $\Lambda$CDM parameter $\Omega_{\Lambda 0}$, so $\Xi$ plays the role of a cosmological constant. $H_0$ is, of course, the current value of the Hubble-Lemaître parameter, and $G_{4,0}=G_4(\varphi_0)$, where $\varphi_0$ is the current background value of the scalar field. Eqs. (\ref{H(a) analitico}) coincides with the results already obtained in our previous work \cite{Gervani:2026zqm}, while Eq. (\ref{G4 analitico}) is a generalized version of our previous result. 

We note that the background evolution of the boost factor, $G_4(\varphi)$, in gCRG coincides with the background evolution of the scalar field in CRG. Therefore, the behavior of $G_4$ will be exactly equal to the one illustrated in Figure 1 of \cite{Gervani:2026zqm} for the scalar field: the boost factor decreases sharply as a function of the scale factor. This result, which turned out to be a pivotal criticality for the Brans-Dicke theory previously analysed, could potentially prove determinant also for the more general case of Horndeski theories. In the next section, we derive the linear perturbation theory for gCRG, in order to check if this is the case. 

\section{Linear perturbation theory and growth factor equation in gCRG}
\label{lin pert th}
In this section we develop the linear perturbation theory associated with our proposed Horndeski model, with the aim to apply such a theory to the study of structure formation in a gCRG universe. Scalar and tensor perturbations to the metric can be considered separately, since, at first perturbative order, the resulting evolution equations do not contain terms that mix the two kinds of perturbations. The scalar perturbations describe the formation of the large-scale cosmic structure, whereas the tensor perturbations are potentially useful to study the propagation of gravitational waves.

We remark that in the model under scrutiny there is no dark matter nor dark energy, since both of these effects should be accounted for by the single scalar field introduced in the Horndeski theory. Our approach is similar but substantially different from that of the so-called Unified Dark Matter (UDM) models, in which the scalar field acts as an effective dark fluid, that, by clustering, provides the gravitational potential wells where the baryonic matter falls \cite{Bertacca:2008uf, Bertacca:2010ct}. In our case there is no clustering of the scalar field, but the  clustering of baryonic matter is expected to be driven by the boosted gravitational force, as computed in Eq. (\ref{RG equation}). Therefore, it is imperative that the boost factor, which, in the case of a Horndeski theory, corresponds to the inverse of the function $G_4(\varphi)$, behaves in a way that allows for structure growth. Thus, in order to study how the boost factor can impact structure formation, we employ linear perturbation theory. First, we derive the perturbed versions of the Einstein field equations, the KG equation, and the continuity and Euler equations. Then, we employ this set of equations to derive the equation for the evolution of the growth factor in a Horndeski universe. 

We introduce scalar perturbations on a flat FLRW background metric  in the Newtonian gauge
\begin{equation}
    \mathrm{d} s^{2}=a^{2}(\eta)\left[-(1+2 \Psi) \mathrm{d} \eta^{2}+(1+2 \Phi) \delta_{i j} \mathrm{d} x^{i} \mathrm{d} x^{j}\right]\,,
    \label{perturbed FLRW metric}
\end{equation}
where $\eta$ is the conformal time given by $ \mathrm{d} \eta = \mathrm{d}t/a(t)$, and $\Psi$ and $\Phi$ are the two independent gravitational potentials. Following steps similar to those of our previous work \cite{Gervani:2026zqm}, we can obtain the growth factor equation in the case of the Horndeski theory under study (for a detailed derivation, see Appendix \ref{Linear perturbation theory in gCRG})
\begin{align*}
         & \delta^{\prime\prime} + \left(2 + \frac{H^\prime}{H}\right)\delta^\prime - \frac{8\pi G}{G_4}\frac{\rho_{m0}}{H^2a^3}\delta - \frac{\nabla^2\delta\varphi}{a^2H^2}\frac{G_{4\varphi}}{G_4} + 2\frac{\Xi}{H^2}\delta\varphi\frac{G_{4\varphi}}{G_4}  = -6\left(1+\frac{H^\prime}{H}\right)\Psi\\
         & - 3\Psi^\prime  + 3\frac{G_{4\varphi}}{G_4}\left(1+\frac{H^\prime}{H}\right)\delta\varphi + \frac{G_{4\varphi}}{G_4}\left( 6\Psi\varphi^\prime - 3\Phi^\prime\varphi^\prime - 3\delta\varphi^\prime \right) - 3\frac{G_{4\varphi\varphi}}{G_4}\varphi^\prime\delta\varphi \\
         & - \left( \frac{2G_{4\varphi}G_{4\varphi\varphi}}{G_4^2}  - \frac{G_{4\varphi}^3}{G_4^3} \right)\left(\varphi^\prime\right)^2\delta\varphi - 2\left(\frac{G_{4\varphi}^2}{G_4^2} - \frac{G_{4\varphi\varphi}}{G_4}\right)\varphi^\prime\delta\varphi^\prime \, . \numberthis
     \label{delta evolution}
\end{align*}
In this equation $\delta=\delta\rho/\rho$ is the density contrast of matter perturbations, and a prime indicates derivation with respect to the time variable $N=\ln a$. This equation for the evolution of the density contrast is coupled to the perturbed KG equation
\begin{align*}
    &\delta \varphi^{\prime\prime} + \left(3+\frac{H^\prime}{H}\right)\delta\varphi^\prime - \frac{4\pi G}{G_{4\varphi}}\frac{\rho_{m0}}{H^2a^3}\delta - \frac{\nabla^{2} \delta\varphi}{a^2H^2} + 2\frac{\Xi}{H^2}\delta\varphi = 2 \Psi \varphi^{\prime\prime} + 2\Psi\left(1+\frac{H^\prime}{H}\right)\varphi^\prime \\
    & + 4\varphi^\prime\Psi+ \varphi^\prime\left(\Psi^\prime - 3\Phi^\prime\right)
    - \frac{G_{4\varphi\varphi}}{G_{4\varphi}}\delta\varphi\left[\varphi^{\prime\prime}+\left(3+\frac{H^\prime}{H}\right)\varphi^\prime\right] - \frac{G_{4\varphi\varphi\varphi}}{G_{4\varphi}}\left(\varphi^\prime\right)^2\delta\varphi - 2\frac{G_{4\varphi\varphi}}{G_{4\varphi}}\varphi^\prime\delta\varphi^\prime \\
    & + 2\frac{G_{4\varphi\varphi}}{G_{4\varphi}}\Psi(\varphi^\prime)^2 \,, \numberthis
    \label{perturbed KG}
\end{align*}
which describes the evolution of the scalar field perturbations, $\delta\varphi$, and to the background KG equation, which describes the evolution of the background value of the scalar field, $\varphi$:
\begin{equation}
    \varphi^{\prime\prime} + \left(3 + \frac{H^\prime}{H}\right)\varphi^\prime - \frac{4\pi G}{G_{4\varphi}}\frac{\rho_{m0}}{H^2a^3} + 2\frac{\Xi}{H^2}\frac{G_4}{G_{4\varphi}} = \frac{G_{4\varphi\varphi}}{G_{4\varphi}}\left(\varphi^\prime\right)^2.
    \label{background KG}
\end{equation}

These three equations are the generalizations of the growth factor and KG equations we obtained in \cite{Gervani:2026zqm}, specifically Eq. (4.9), Eq. (4.5e) and Eq. (3.13), respectively. In particular, the source term in Eq. (\ref{delta evolution}), that in the case of the Brans--Dicke theory appeared divided by the scalar field, is now divided by the function $G_4$. This similarity is expected, given our interpretation of $G_4$ as the gravitational boost factor. We already saw, in Section \ref{bkg}, how the background evolution of the boost factor in our Horndeski theory corresponds to the evolution of the scalar field in the Brans--Dicke case. Here, again, the role formerly played by $\varphi$ is assumed by $G_4(\varphi)$. In \cite{Gervani:2026zqm} we have shown how, due to the background evolution of $\varphi$ and its presence in the denominator of the source term of the grow factor equation, structure formation was hindered. Now, we find ourselves in a position to extend the  results of \cite{Gervani:2026zqm} to the larger class of Horndeski theories: since $G_4$ evolves exactly as $\varphi$ at the background level and it also appears in the denominator of the source term, we can expect that the qualitative behavior of the density contrast will be very similar to the Brans-Dicke theory. Indeed, in the following section we will numerically solve the growth factor equation and confirm these expectations. Details of the necessary calculations conducted for the numerical solution are reported in Appendix \ref{numerical solution}.

\section{QSA numerical solution}
\label{QSA numerical solution}
As discussed in Appendix \ref{Linear perturbation theory in gCRG}, it is difficult to solve the whole system of coupled differential equations in Eqs. (\ref{delta evolution}), (\ref{perturbed KG}) and (\ref{background KG}), because there is no physically-motivated way to set the initial conditions on the scalar field and its derivatives. This situation is different from the one previously considered in \cite{Gervani:2026zqm}, because in that case an analytical solution for $\varphi(a)$ was known and therefore the values of $\varphi$ and $\varphi^\prime$ at $z=10^{-3}$ were chosen as initial conditions. Fortunately, as was already argued in \cite{Gervani:2026zqm}, we can resort to the quasi-static approximation (QSA) to understand the qualitative behavior of Eq. (\ref{delta evolution}) and provide a more straightforward way to integrate it. 

In QSA \cite{Silvestri2013,DeFelice2012,Pace2019,Pace2021}, it is assumed that the time derivatives of the metric scalar perturbations can be neglected, because they are of the order of the Hubble-Lemaître parameter. Such terms are negligible on sufficiently small scales, i.e. sub-horizon scales. If we neglect these terms, our set of equations considerably simplifies, and we can find a solution for the evolution of the density contrast $\delta$ via a simplified numerical approach, one that does not require to solve for the evolution of the scalar field and its perturbations. 

The basic approach of QSA is the following: one can write three Poisson-like equations, which relate the potentials $\Psi$ and $\Phi$ and the scalar field $\varphi$ to the matter content of the universe, which, in this case, is only given by the baryonic contribution. The Newtonian gravitational constant $G$ appearing in these equations will not be a constant, but a generic function of time and scale that can be expressed in terms of the following functions (we refer the reader to Eqs.~(B.1)-(B.5) of~\cite{Pace2019} for the complete definitions):
\begin{subequations}
\begin{align}
 M^2 = &\, 2G_4\,, \\
 \alpha_M = &\, \frac{1}{HM^2}\frac{\mathrm{d}M^2}{\mathrm{d}t} = \frac{\dot{G_4}}{HG_4}\,,\\
 \alpha_K = &\, \frac{1}{M^2H^2}2X\frac{G_{4\varphi}^2}{G_4} = -\frac{\dot{G_4}^2}{H^2G_4^2}\,,\\
 \alpha_B = &\, \frac{1}{2M^2H}\frac{\mathrm{d}M^2}{\mathrm{d}t} = \frac{1}{2}\alpha_M\,,\\
 \alpha_T = &\, \alpha_H = 0\,.
\end{align}
\end{subequations}
Here, we consider the two effective gravitational constants $\mu_{Y}$ and $\mu_{Z}$, defined by the modified Poisson equations
\begin{equation}
 \nabla^2\Psi = 4\pi G\mu_Y a^2 \delta\rho_{\rm m}\,, \quad 
 \nabla^2\Phi = -4\pi G\mu_Z a^2 \delta\rho_{\rm m}\,.
\end{equation}
One can also define the combination $\mu_W = \Sigma = (\mu_{Y}-\mu_{Z})/2$, that enters as a modification of the standard weak gravitational lensing effect. Only $\mu_{Y}$ and $\mu_W$ are directly observable, whereas no quantity which depends on $\mu_{Z}$ alone is known. In fact, in the context of structure formation, $\mu_Y$ is responsible for observable deviations from the predictions of GR. This is evident from Eq. (\ref{passaggio growth factor}), which only contains the Laplacian of the potential $\Psi$. Following \cite{Pace2021}, we can compute $\mu_Y$ as
\begin{equation}
 \mu_Y= \,\mu_{Y,\infty} = \, \frac{\alpha c_{\delta\phi}^2(1+\alpha_T) + 2[\alpha_B(1+\alpha_T) + \alpha_T - \alpha_M]^2}{\alpha c_{\delta\phi}^2 \bar{M}^2}\,, 
  \label{eq:muy}
\end{equation}
 where $\alpha=\alpha_K + 6\alpha_B^2$ and the sound speed of the scalar field perturbations, $c_{s, \delta\varphi}^2$, is \cite{Pace2021}
\begin{equation}
	c_{s, \delta\varphi}^2 = - \frac{2\left(1+\alpha_B\right)\left[\alpha_B\left(1+\alpha_T\right) - \left(\alpha_M - \alpha_T\right) + \frac{\dot{H}}{H^2}\right] + 2\dot{\alpha_B}H^{-1}}{\alpha} - \frac{\rho_{\rm m} + P_{\rm m}}{\alpha M^2H^2} = 1\, .
	\label{cs2}
\end{equation}
 Inserting the above relations into Eq. (\ref{eq:muy}) yields $\mu_Y = 1/G_4$. This result is consistent with the interpretation of $G_4$ as the boost factor in our modified theory of gravity, and also agrees with the same result obtained for the peculiar case of the Brans--Dicke theory in \cite{Gervani:2026zqm}. 

\begin{figure}[htbp]
        \includegraphics[width=\textwidth]{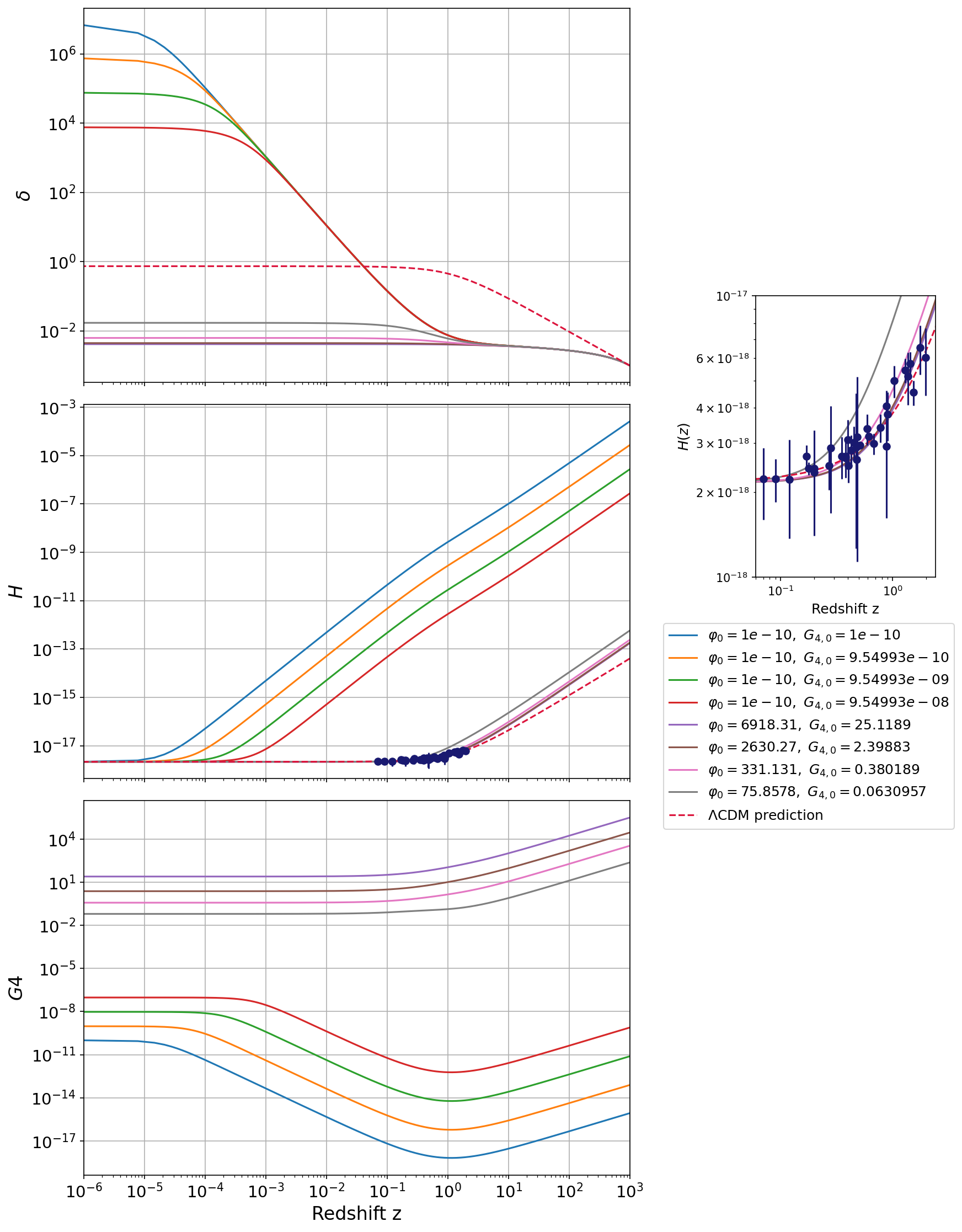}
    \caption{Evolution of the density contrast $\delta$ (top panel), of the Hubble-Lemaître parameter $H$ (middle panel) and of the boost factor function $G_4$ (bottom panel) as a function of redshift. In the middle panel, the solid points with error bars show the observational estimates from~\cite{Yu:2017iju}. The smaller panel on the right depicts a zoomed-in region of the $H$ plot. Different curves assume different values for the free parameters $\varphi_0$ and $G_{4,0}$; the other free parameter, $\Omega_{\Xi 0}$, is fixed to the upper limit of Eq.~(\ref{omega_xi condition 2}), following the results of \cite{Gervani:2026zqm} and \cite{Sanna}. The dashed red line in the top and middle panels represents the $\Lambda$CDM prediction.}
    \label{delta evolution figure}
\end{figure}

Knowing $\mu_Y$ and neglecting the terms containing the time derivatives of the scalar potentials, Eq. (\ref{passaggio growth factor}) reduces to 
\begin{equation}
         \delta^{\prime\prime} + \left(2 + \frac{H^\prime}{H}\right)\delta^\prime - \frac{4\pi G}{G_4}\frac{\rho_{m0}}{H^2a^3}\delta = 0.
         \label{gf QSA}
\end{equation}
This simplified differential equation can be solved without assuming any specific initial condition on the scalar field. We note here that this equation can also be obtained starting from the more general results obtained in Appendix \ref{numerical solution}: in fact, combining Eq. (\ref{equation for delta evolution}) with Eq. (\ref{equation for delta phi evolution}) multiplied by $G_{4\varphi}/G_4$, we obtain
\begin{align*}
   & \delta^{\prime\prime} + \left(2 + \frac{H^\prime}{H}\right)\delta^\prime - \frac{4\pi G}{G_4}\frac{\rho_{m0}}{H^2a^3}\delta - \frac{G_{4\varphi}}{G_4} \delta \varphi^{\prime\prime} - \frac{H^\prime}{H}\frac{G_{4\varphi}}{G_4}\delta\varphi^\prime + 3\frac{G_{4\varphi}}{G_4}\left(1+\frac{H^\prime}{H}\right)\delta\varphi = \\
   & \frac{G_{4\varphi\varphi}}{G_{4}}\delta\varphi\left(\varphi^{\prime\prime}+\frac{H^\prime}{H}\varphi^\prime\right) + \left(\frac{G_{4\varphi\varphi\varphi}}{G_{4}} - \frac{2G_{4\varphi}G_{4\varphi\varphi}}{G_4^2} + \frac{G_{4\varphi}^3}{G_4^3}\right)\left(\varphi^\prime\right)^2\delta\varphi - 2\left(\frac{G_{4\varphi}^2}{G_4^2} - \frac{G_{4\varphi\varphi}}{G_{4}}\right)\varphi^\prime\delta\varphi^\prime \, .\numberthis
\end{align*}
However, we can neglect all the terms containing $\delta\varphi$ in this equation, because, on sub-horizon scales, they are sub-leading with respect to the source term: from Eq. (\ref{equation for delta phi evolution}) it is clear that the source term is $\propto k^2\delta\varphi$, while all the other terms are proportional to $k\delta\varphi$, $\delta\varphi$, $H\delta\varphi$ or $H^2\delta\varphi$. Neglecting such terms leaves us with Eq. (\ref{gf QSA}). The result of the integration of Eq. (\ref{gf QSA}) is depicted in the first panel of Figure \ref{delta evolution figure}. The other two panels depict the analytical solutions for $H$, Eq. (\ref{H(a) analitico}), and $G_4$, Eq. (\ref{G4 analitico}).

In Figure \ref{delta evolution figure}, different curves represent different choices for the free parameters $(\varphi_0, G_{4,0})$ of the theory. Following the discussion of our previous work \cite{Gervani:2026zqm}, the parameters $\rho_{\text{m}0}$ and $H_0$ [appearing in Eq.(\ref{H(a) analitico})] are considered known and are taken to be equal to their current $\Lambda$CDM estimates. Indeed, the estimation of the baryonic density $\rho_{\text{m}0}$ depends on the assumed underlying cosmological model. Therefore, in principle, we should estimate this parameter from the data by assuming the Horndeski model under analysis. The impact of this simplifying assumption remains to be estimated. Since $\rho_{\text{m}0}$ and $\Omega_{\text{m}0}$ refer to the baryonic component only, we adopt $\Omega_{\text{m}0} = 0.05$~\cite{Planck:2018vyg}. We also adopt $H_0 = 67$ km/s/Mpc. 

On the other hand, the parameters $\Omega_{\Xi 0}$, $\varphi_0$ and $G_{4,0}$ are unknown. Other a priori unknown quantities, such as $\Xi$ and $\Omega_0$, depend on these parameters through the definitions of Eqs.~(\ref{omega, omega_xi}). Also, the parameters $\Omega_{\Xi 0}$ and $\Omega_0$ must satisfy the condition (\ref{omega_xi condition}), which can be recast as
\begin{equation}
    \Omega_0 > 0 \, , \qquad 0 < \Omega_{\Xi 0} \leq \frac{\Omega_0 + 1}{2}\,.
    \label{omega_xi condition 2}
\end{equation}
However, it has been shown that the values of $\Omega_{\Xi 0}$ that saturate the upper bound in Eq. (\ref{omega_xi condition 2}) provide the best-fit for measurements of the Hubble-Lemaître parameter coming from both a SNe dataset \cite{Sanna} and a cosmic chronometers dataset \cite{Gervani:2026zqm, M_Moresco_2012}. Since the value of $\Omega_0$ is related to that of $G_{4,0}$ through Eq. (\ref{omega, omega_xi}), once we have chosen $G_{4,0}$ we decide to fix $\Omega_{\Xi 0}$ to be equal to $(\Omega_0+1)/2$. With this choice, our Horndeski theory is completely characterized by the choice of the pair of parameters $(\varphi_0, G_{4,0})$, and Eq. (\ref{delta evolution}) can be numerically integrated. Initial conditions on $\delta$ have been set following the prescription $\delta \propto a$, and therefore $\delta(z=1000) = 10^{-3}$.

The second panel of Figure \ref{delta evolution figure} also reports several data points for the Hubble-Lemaître parameter as a function of redshift. The smaller zoomed-in panel on the right provides a better visual inspection of the data points and their error bars. In the literature, there exist two main data sets for $H(z)$~\cite{Hu:2025fsz, Hu:2024big, CosmoVerse:2025txj}:  the cosmic chronometer data set~\cite{Moresco:2022phi, Moresco:2023zys}, and the baryonic acoustic oscillations (BAO) dataset~\cite{Yang:2024kdo}. As already done in \cite{Gervani:2026zqm}, we decide to employ the values reported in~\cite{Yu:2017iju} and only use data points based on cosmic chronometers. Our choice is motivated as follows: the cosmic chronometer approach, contrary to the BAO measurement, is not dependent on the choice of a specific cosmological model, because it estimates the age of a galaxy based on evolution models of stellar populations~\cite{Yu:2017iju, M_Moresco_2012, Zhang_2014}. Estimates based on the BAO scale are computed by solving the Boltzmann equations, an operation that must necessarily assume a cosmological model (usually $\Lambda$CDM).

In Figure \ref{delta evolution figure}, two families of solutions are represented: those that maximize growth (blue, orange, green and red curves) and those that successfully fit the background evolution of the Hubble-Lemaître parameter (grey, pink, brown and purple curves). One can immediately see that there is no overlap between the two families: a given choice of free parameters can either provide a substantial growth of the cosmic structure in linear perturbation theory, or it can properly describe the background expansion rate. 

Solutions that provide a considerable growth (i.e. those that reach at least, $\delta\sim1$, a limit beyond which the depicted evolution loses its significance since perturbation theory breaks down) are those with values of $G_{4,0}$ well below unity. The lower the value of $G_{4,0}$, the higher the final achieved growth, with the highest values of $\delta$ reached at the bottom of the explored range, for $G_{4,0}=10^{-10}$. This result is sensible, considering that the gravity boost factor is $G_4^{-1}$. At the same time, we see that the resulting growth is fairly insensitive to the value of $\varphi_0$: we plot curves obtained with $\varphi_0=10^{-10}$, but choosing any other value does not change the results significantly.

On the other hand, the solutions that can fit the $H(z)$ datapoints require values of $G_{4,0}$ that are much higher than those of the previous family. Such high values of $G_{4,0}$ do not provide a sufficiently strong gravity boost, resulting in a suppression of the growth of the density contrast. If one tries to lower $G_{4,0}$ below unity (for instance, in the case of the grey curve) the $H(z)$ fit becomes poor before one can achieve an appreciable growth, as it is clear by comparing the first panel and the zoomed-in smaller panel. Therefore, there seems to exist an inescapable trade-off effect between structure growth and background expansion for the value of the parameter $G_{4,0}$. 

Even disregarding the problem so far illustrated, the growth predicted by the family of curves that maximize $\delta$ is still, in itself, problematic, because the growth predicted is very delayed, with the density contrast that reaches the non-linear regime only at redshifts $z<1$. This prediction is incompatible with recent observations of very ancient galaxies at redshifts as high as $z\sim 14$~\cite{naidu2025cosmicmiracleremarkablyluminous}. The $\Lambda$CDM prediction for the growth factor is much higher than the gCRG one at larger redshifts, even if also in this case there is some tension between the predicted number density of massive galaxies and the observed one~\cite{Boylan-Kolchin:2011qkt, Labbe_2023}.  

We can also pinpoint precisely, in our equations, the origin of this delay in structure formation: it is due to the coefficient of the source term in Eq.~(\ref{delta evolution}): $8\pi G\rho_{{\rm m}0}/(G_4H^2a^3)$. Given the form of $G_4$ in Eq.~(\ref{G_4 analitico}), one can easily see that this coefficient turns out to be proportional to $H^{-1}$. The second panel of Figure \ref{delta evolution figure} shows that the Hubble-Lemaître parameter decreases with $a^{-\sqrt{3}}$ [Eq.~(\ref{H(a) analitico})]. Therefore, the contribution of the source term decreases with increasing redshifts and is negligible at high redshift: the evolution of the perturbations is only triggered at relatively small redshifts, when this term becomes appreciable. We recall that in the $\Lambda$CDM model, this coefficient is almost a constant. 

Our numerical analysis has confirmed the expectations illustrated at the end of Section \ref{lin pert th}. We have thus extended the results of our previous work \cite{Gervani:2026zqm} to the broader class of Horndeski theories considered in this work. 

\section{Horndeski theory in a curved universe}
\label{Horndeski theory in a curved universe}
As a further test and possible generalization of our original Brans-Dicke approach in \cite{Gervani:2026zqm}, we considered the possibility that the space curvature of the universe might influence  structure formation. We start from the metric for a universe with curvature $\mathcal{K}$, which is homogeneous and isotropic in the background but contains scalar perturbations written in the Newtonian gauge: in Cartesian coordinates, this metric is
\begin{equation}
     {\rm d}s^2 = -a^2(1+2\Psi){\rm d}\eta^2 + a^2\frac{(1+2\Phi)}{(1+\mathcal{K} r^2/4)^2}\delta_{ij}{\rm d}x^i{\rm d}x^j
    \label{curved perturbed metric}
\end{equation}
where $r$ is the Euclidean distance to the origin of coordinates, $r^2=x_i x^i$. In Appendix \ref{Linear perturbation theory in a Horndeski curved universe}, we derive the equations for the evolution of linear perturbations in a general Horndeski model, and we later specialize it to the theory defined by the choices in Eqs. (\ref{choices for RG in Horndeski}). This calculation yields the following differential equation, that describes the evolution of the growth factor in a curved universe:
\begin{align*}
         & \delta^{\prime\prime} + \left(2 + \frac{H^\prime}{H}\right)\delta^\prime - \frac{8\pi G}{G_4}\frac{\rho_{m0}}{H^2a^3}\delta - \frac{\nabla^2\delta\varphi}{\gamma^2 a^2H^2}\frac{G_{4\varphi}}{G_4} + \frac{\mathcal{K}}{2\gamma}\frac{x^j\partial_j\delta\varphi}{a^2H^2}\frac{G_{4\varphi}}{G_4} + 2\frac{\Xi}{H^2}\delta\varphi\frac{G_{4\varphi}}{G_4} \\
         &= -6\left(1+\frac{H^\prime}{H}\right)\Psi - 3\Psi^\prime  + 3\frac{G_{4\varphi}}{G_4}\left(1+\frac{H^\prime}{H}\right)\delta\varphi + \frac{G_{4\varphi}}{G_4}\left( 6\Psi\varphi^\prime - 3\Phi^\prime\varphi^\prime - 3\delta\varphi^\prime \right) \\
         & - 3\frac{G_{4\varphi\varphi}}{G_4}\varphi^\prime\delta\varphi - \left( \frac{2G_{4\varphi}G_{4\varphi\varphi}}{G_4^2}  - \frac{G_{4\varphi}^3}{G_4^3} \right)\left(\varphi^\prime\right)^2\delta\varphi - 2\left(\frac{G_{4\varphi}^2}{G_4^2} - \frac{G_{4\varphi\varphi}}{G_4}\right)\varphi^\prime\delta\varphi^\prime \, . \numberthis
     \label{delta curved evolution}
\end{align*}
This equation is identical to its counterpart, Eq. (\ref{delta evolution}), in a flat universe $\mathcal{K}=0$, except for the fact that the Laplacian operator has been substituted, in Eq. (\ref{delta curved evolution}), with its curved equivalent, $\nabla^2\xrightarrow[]{}\nabla^2/\gamma^2 - (\mathcal{K}/2\gamma)x^j\partial_j$. 

We can exclude the possibility that such a small modification of the equation could lead to a significant change in the behavior of the growth factor. In fact, as it was shown in \cite{Gervani:2026zqm}, the terms involving the perturbations of the scalar field turn out to be negligible when compared with the other terms of Eq. (\ref{delta curved evolution}). Moreover, the new term containing the curvature, $(\mathcal{K} x^j\partial_j\delta\varphi G_{4\varphi})/(2\gamma a^2H^2 G_4)$, is negligible with respect to the Laplacian term when considered on scales smaller than the cosmological horizon. This condition is easily seen through a Fourier transformation: while the Laplacian term is $\propto k^2$, where $k$ is the Fourier scale considered, the new curvature term is only $\propto k$, and thus smaller than the Laplacian term on small scales. For all these reasons, we expect that the curvature cannot yield  significant deviations from the results of the previous Section. 

Another way to see that no significant deviation is expected is to proceed as in Section \ref{QSA numerical solution}: since the specific Horndeski model is not modified by the presence of curvature, we obtain again Eq. (\ref{gf QSA}) under the QSA assumption. The difference in the Laplacian operator is neglected because the QSA is only valid on sub-horizon scales, as discussed above.

The study of QSA can also be performed for a general Horndeski theory with spatial curvature using the field equations presented in \cite{Gambino2024}. Taking their Eqs.~(8)--(12) written in configuration space, to apply the QSA we move to the Fourier space. However, it is not possible to perform a simple Fourier transform, because the covariant derivative $D^2$ does not become $-k^2$, but a differential operator in momentum space. Therefore, we expand our functions in the harmonics $\chi(\vec{x}) = \sum\chi_qQ_{lmn}$ such that $D^2 Q_{lmn} = -(q^2-3\mathcal{K})Q_{lmn}$ (see Appendix \ref{QSA in a Horndeski curved universe} for details). Thus, terms like $D^2 + 3\mathcal{K}$ simply become $-q^2$, and the field equations in the presence of spatial curvature have the following modifications with respect to the flat case: the terms with $\mathrm{K} \equiv k/aH$ are now written as $Q = q/aH$; moreover, there are two additional terms proportional to $\mathcal{K}$ in the equations for the trace of the (ij) component and in the equations of motion for $\pi$. Details of these calculations can be found in Appendix \ref{QSA in a Horndeski curved universe}. The application of QSA implies that $Q\gg 1$ and that we neglect the time derivatives of the potential and of the scalar field. Furthermore, for consistency, we need to neglect the terms in which the spatial curvature $\mathcal{K}$ appears. This quantity represents the background curvature and, from the Friedmann equation, where $M_{\rm pl}= (8\pi G)^{-1/2}$,
\begin{equation}
    H^2 + \frac{\mathcal{K}}{a^2} = \frac{1}{3M_{\rm pl}^2}\left(\rho_{\rm m} + \rho_{\rm de}\right)\,,
\end{equation}
this quantity is of order $H^2$, exactly as the terms neglected. Therefore, the final equations are the same as in \cite{Pace2021}, with the replacement $\mathrm{K}^2 \to \mathrm{Q}^2$. Hence, we can see that the spatial curvature does not substantially affect the growth of structures.

\section{Conclusions}
\label{Conclusions}
In this work we generalized our previous study of a Brans–Dicke theory of gravity without dark matter \cite{Gervani:2026zqm} to the full class of Horndeski theories, reducing to Refracted Gravity (RG) in the weak-field limit. Our goal was determining whether a broader scalar-tensor framework could succeed where the specific Brans–Dicke realization had failed in reproducing the observed growth of cosmic structure. We began by deriving, in Section \ref{Horndeski theory and generalized Covariant Refracted Gravity} and Appendix \ref{Weak field limit of a Horndeski theory}, the general conditions on the Horndeski functions $G_2$, $G_3$, and $G_4$ required for the theory to reduce, in its non-relativistic weak-field limit, to RG. We found that this reduction is possible for a wide family of theories, summarized in Eqs. (\ref{choices for RG in Horndeski}), in which the function $G_4(\phi)$ plays the role of the gravitational permittivity $\epsilon(\phi)$ of RG for essentially any functional dependence on the scalar field, rather than only the linear relation assumed in our earlier Brans–Dicke analysis. We referred to this broader theory as generalized Covariant Refracted Gravity (gCRG).

Studying the background evolution of a flat, matter-dominated universe in gCRG (Section \ref{bkg} and Appendix \ref{Background evolution equations in gCRG}), we showed that the boost factor $G_4[\phi(a)]$ obeys the same evolution equation as the scalar field did in our previous Brans–Dicke model, and that the resulting expression for the Hubble–Lemaître parameter $H(a)$ is formally identical to the one we obtained in \cite{Gervani:2026zqm}. This result is a direct consequence of the structural similarity imposed by the RG-matching conditions, and it implies that any success or failure of the background cosmology found previously carries over to the more general Horndeski setting.

The central contribution of this work is the derivation of the linear perturbation theory and the growth factor equation for gCRG, both in a flat universe (Section \ref{lin pert th} and Appendix \ref{Linear perturbation theory in gCRG}) and in a universe with non-zero spatial curvature (Section \ref{Horndeski theory in a curved universe} and Appendix \ref{Linear perturbation theory in a Horndeski curved universe}). We showed that curvature effects introduce no qualitatively new behavior in the regime relevant for structure formation, since the additional curvature terms in the perturbed field equations scale linearly with the Fourier wavenumber $k$, and are therefore negligible compared to the $k^2$ terms that dominate on sub-horizon scales. Employing the quasi-static approximation, which circumvents the difficulty of specifying physically motivated initial conditions for the scalar field perturbations, we numerically integrated the growth factor equation for a range of values of the free parameters $\phi_0$, $G_{4,0}$ and $\Omega_{\Xi 0}$ (Section \ref{QSA numerical solution}).

Our numerical results, summarized in Figure \ref{delta evolution figure}, show a clear tension within the model: parameter choices that correctly reproduce the observed evolution of $H(z)$ require values of $G_{4,0}$ of order unity or larger, which suppress the gravitational boost and delay the growth of density perturbations to the point that no appreciable structure forms until very recent epochs. Conversely, parameter choices that allow $\delta$ to grow into the non-linear regime require $G_{4,0}\ll1$, but such small values fail to reproduce the background $H(z)$ data. No choice of parameters achieves both results simultaneously, and even in the most favorable "growth-maximizing" branch, the density contrast becomes non-linear only at $z<1$ — a prediction incompatible with the existence of massive, evolved galaxies observed at redshifts as high as $z\sim14$ \cite{naidu2025cosmicmiracleremarkablyluminous}. 

We traced the reason of this behavior directly to the structure of the source term of the growth factor equation, Eq. (\ref{delta evolution}), where the standard $4\pi G\rho_m$ term is divided by the boost factor $G_4$. Because $G_4(a)$ decreases sharply toward early times — mirroring the analogous decrease of the scalar field found in the Brans–Dicke case \cite{Gervani:2026zqm} — the source term is strongly suppressed at high redshift, delaying the onset of gravitational instability well past the epoch required by observations. Since this feature is not tied to any particular choice of the potential or kinetic function, but it follows generically from the conditions in Eqs.  (\ref{conditions}) that any Horndeski theory must satisfy in order to reduce to RG in the weak-field limit, we conclude that the delayed structure formation obtained in our earlier Brans–Dicke study is not an artifact of that specific model. Rather, it appears to be a structural obstruction shared by the entire class of Horndeski theories capable of reproducing RG phenomenology on galactic scales.

This conclusion suggests that reconciling a scalar-tensor ``gravity-only'' explanation of both dark matter and dark energy phenomenology with the observed timeline of structure formation might require relaxing one of the assumptions underlying the weak-field correspondence with RG derived here. We leave the exploration of this possibility to future work.

\acknowledgments
We sincerely thank Lorenzo Fatibene for enlightening discussions. We acknowledge partial support from the grant InDark of the Italian National Institute of Nuclear Physics (INFN). This research has made use of NASA’s Astrophysics Data System Bibliographic Services.

\newpage
\appendix

\section{Weak field limit of a Horndeski theory}
\label{Weak field limit of a Horndeski theory}
In this appendix, we derive the conditions, reported in Eq. (\ref{conditions}), under which a generic Horndeski theory reduces to RG in the weak-field limit. We start by deriving the expression for the Ricci scalar: from Eq. (\ref{KG}), we see that  
\begin{align*}
    R & = \frac{1}{G_{4\varphi}}\left\{\nabla^\mu\left[(2G_{2X} - 2G_{3X}\square\varphi - 2G_{3\varphi})\nabla_\mu\varphi + 2G_{3X}\nabla_\mu X\right] - G_{2\varphi} + \nabla_\lambda G_{3\varphi}\nabla^\lambda\varphi\right\}\\
    &= \frac{2 G_{2X} \Delta\varphi}{G_{4\varphi}}
    - \frac{4 G_{2X} \Delta\varphi\,\Phi}{G_{4\varphi}}
    - \frac{2 G_{3X} (\Delta\varphi)^2}{G_{4\varphi}}
    + \frac{8 G_{3X} (\Delta\varphi)^2 \Phi}{G_{4\varphi}}
    + \frac{2 G_{3X} \Delta\Phi\,(\nabla\varphi)^2}{G_{4\varphi}}
    - \frac{G_{2\varphi}}{G_{4\varphi}} \nonumber\\
    &- \frac{G_{3\varphi} \Delta\varphi}{G_{4\varphi}}
    + \frac{2 G_{3\varphi} \Delta\varphi\,\Phi}{G_{4\varphi}}
    + \frac{2 G_{2\varphi X} (\nabla\varphi)^2}{G_{4\varphi}}
    - \frac{4 G_{2\varphi X} \Phi\,(\nabla\varphi)^2}{G_{4\varphi}}
    - \frac{2 G_{3\varphi X} \Delta\varphi\,(\nabla\varphi)^2}{G_{4\varphi}} \nonumber\\
    &+ \frac{8 G_{3\varphi X} \Delta\varphi\,\Phi\,(\nabla\varphi)^2}{G_{4\varphi}}
    - \frac{G_{3\varphi\varphi} (\nabla\varphi)^2}{G_{4\varphi}}
    + \frac{2 G_{3\varphi\varphi} \Phi\,(\nabla\varphi)^2}{G_{4\varphi}}
    - \frac{2 G_{3X} \varphi_{ij}\varphi_{ij}}{G_{4\varphi}}
    + \frac{8 G_{3X} \Phi\,\varphi_{ij}\varphi_{ij}}{G_{4\varphi}} \nonumber\\
    &- \frac{4 G_{3X} \varphi_{iij}\,\varphi_j}{G_{4\varphi}}
    + \frac{16 G_{3X} \Phi\,\varphi_{iij}\,\varphi_j}{G_{4\varphi}}
    + \frac{4 G_{2XX} \varphi_i\,\varphi_{ij}\,\varphi_j}{G_{4\varphi}}
    - \frac{16 G_{2XX} \Phi\,\varphi_i\,\varphi_{ij}\,\varphi_j}{G_{4\varphi}} \nonumber\\
    &- \frac{4 G_{3XX} \Delta\varphi\,\varphi_i\,\varphi_{ij}\,\varphi_j}{G_{4\varphi}}
    + \frac{24 G_{3XX} \Delta\varphi\,\Phi\,\varphi_i\,\varphi_{ij}\,\varphi_j}{G_{4\varphi}}
    - \frac{4 G_{3\varphi X} \varphi_i\,\varphi_{ij}\,\varphi_j}{G_{4\varphi}}
    + \frac{16 G_{3\varphi X} \Phi\,\varphi_i\,\varphi_{ij}\,\varphi_j}{G_{4\varphi}} \nonumber\\
    &- \frac{4 G_{3XX} \varphi_{ij}\,\varphi_{ik}\,\varphi_j\,\varphi_k}{G_{4\varphi}}
    + \frac{24 G_{3XX} \Phi\,\varphi_{ij}\,\varphi_{ik}\,\varphi_j\,\varphi_k}{G_{4\varphi}}
    - \frac{G_{3} \Delta\varphi\,\Psi_i}{G_{4\varphi}\,\varphi_i}
    - \frac{G_{3\varphi} (\nabla\varphi)^2 \Psi_i}{G_{4\varphi}\,\varphi_i} \nonumber\\
    &+ \frac{2 G_{2X} \varphi_i\,\Psi_i}{G_{4\varphi}}
    - \frac{4 G_{3X} \Delta\varphi\,\varphi_i\,\Psi_i}{G_{4\varphi}}
    - \frac{G_{3\varphi} \varphi_i\,\Psi_i}{G_{4\varphi}}
    - \frac{2 G_{3\varphi X} (\nabla\varphi)^2 \varphi_i\,\Psi_i}{G_{4\varphi}}
    - \frac{6 G_{3X} \varphi_{ij}\,\varphi_j\,\Psi_i}{G_{4\varphi}} \nonumber\\
    &- \frac{4 G_{3XX} \varphi_i\,\varphi_j\,\varphi_k\,\varphi_{kj}\,\Psi_i}{G_{4\varphi}}
    + \frac{G_{3} \varphi_j\,\Psi_{iij}}{G_{4\varphi}\,\varphi_i^2}
    - \frac{G_{3} \varphi_j\,\Psi_{ij}}{G_{4\varphi}\,\varphi_i}
    - \frac{2 G_{3X} \varphi_i\,\varphi_j\,\Psi_{ij}}{G_{4\varphi}} \nonumber\\
    &- \frac{G_{3} \Delta\varphi\,\Phi_i}{G_{4\varphi}\,\varphi_i}
    - \frac{G_{3\varphi} (\nabla\varphi)^2 \Phi_i}{G_{4\varphi}\,\varphi_i}
    + \frac{2 G_{2X} \varphi_i\,\Phi_i}{G_{4\varphi}}
    - \frac{4 G_{2XX} (\nabla\varphi)^2 \varphi_i\,\Phi_i}{G_{4\varphi}}
    + \frac{4 G_{3XX} \Delta\varphi\,(\nabla\varphi)^2 \varphi_i\,\Phi_i}{G_{4\varphi}} \nonumber\\
    &- \frac{G_{3\varphi} \varphi_i\,\Phi_i}{G_{4\varphi}}
    + \frac{2 G_{3\varphi X} (\nabla\varphi)^2 \varphi_i\,\Phi_i}{G_{4\varphi}}
    + \frac{8 G_{3XX} (\nabla\varphi)^2 \varphi_i\,\varphi_{ij}\,\Phi_i}{G_{4\varphi}}
    + \frac{2 G_{3X} \varphi_{ij}\,\varphi_j\,\Phi_i}{G_{4\varphi}} \nonumber\\
    &- \frac{4 G_{3XX} \varphi_i\,\varphi_j\,\varphi_k\,\varphi_{kj}\,\Phi_i}{G_{4\varphi}}
    + \frac{G_{3} \varphi_j\,\Phi_{iij}}{G_{4\varphi}\,\varphi_i^2}
    - \frac{G_{3} \varphi_j\,\Phi_{ij}}{G_{4\varphi}\,\varphi_i}
    - \frac{2 G_{3X} \varphi_i\,\varphi_j\,\Phi_{ij}}{G_{4\varphi}}\, . \numberthis
    \label{Ricci scalar}
\end{align*}
In this expression a latin subscript indicates partial differentiation with respect to a Cartesian spatial direction, while $\Delta=\partial_i\partial_i$ and $(\nabla\varphi)^2=\partial_i\varphi\partial_i\varphi$. Repeated indices follow the Einstein summation convention. Notice that in Eq. (\ref{Ricci scalar}) we are already neglecting terms that are second order in the perturbative quantities $\Phi$ and $\Psi$. Substituting this result into the $00$ component of the Einstein field equations (Eqs. \ref{EE}), one gets, again to first order in $\Phi$ and $\Psi$,
\begin{align*}
&   \nabla^2\Psi G_{4} 
    + \frac{\nabla^2\Phi\, (\nabla\varphi)^{2}\, G_{4}\, G_{3X}}{G_{4\varphi}} 
    - \Phi\left[
    -(\nabla\varphi)^{2} G_{3\varphi}
    + \frac{2\,\nabla^2\varphi\, G_{4}\, G_{2X}}{G_{4\varphi}} 
    - \frac{4\,(\nabla^2\varphi)^{2}\, G_{4}\, G_{3X}}{G_{4\varphi}} \right. \\
& \left.   - \frac{\nabla^2\varphi\, G_{4}\, G_{3\varphi}}{G_{4\varphi}}
    - 2\,\nabla^2\varphi\, G_{4\varphi}
    + \frac{2 (\nabla\varphi)^{2} G_{4} G_{2\varphi X}}{G_{4\varphi}}
    - \frac{4\,\nabla^2\varphi\,(\nabla\varphi)^{2}\, G_{4} G_{3\varphi X}}{G_{4\varphi}}
    - \frac{(\nabla\varphi)^{2}\, G_{4} G_{3\varphi\varphi}}{G_{4\varphi}} \right. \\
& \left.   - 2(\nabla\varphi)^{2} G_{4\varphi\varphi}
    - \frac{4 G_{4} G_{3X}\, \varphi_{ij}^{2}}{G_{4\varphi}}
    - \frac{8 G_{4} G_{3X}\, \varphi_{iij}\, \varphi_{j}}{G_{4\varphi}}
    - 4 G_{3X} \varphi_{i}\varphi_{ij}\varphi_{j}
    + \frac{8 G_{4} G_{2XX} \varphi_{i}\varphi_{ij}\varphi_{j}}{G_{4\varphi}} \right. \\
& \left.  - \frac{12 \nabla^2\varphi\, G_{4} G_{3XX} \varphi_{i}\varphi_{ij}\varphi_{j}}{G_{4\varphi}}
    - \frac{8 G_{4} G_{3\varphi X} \varphi_{i}\varphi_{ij}\varphi_{j}}{G_{4\varphi}} 
    - \frac{12 G_{4} G_{3XX} \varphi_{ij}\varphi_{ik}\varphi_{j}\varphi_{k}}{G_{4\varphi}}
    \right] \\
&   + \Psi\left[
    G_{2}
    - (\nabla\varphi)^{2} G_{3\varphi}
    + \frac{2\nabla^2\varphi\, G_{4} G_{2X}}{G_{4\varphi}}
    - \frac{2\nabla^2\varphi^{2}\, G_{4} G_{3X}}{G_{4\varphi}}
    - \frac{2\nabla^2\varphi\, (\nabla\varphi)^{2} G_{4} G_{3X}}{G_{4\varphi}}
    - \frac{G_{4}G_{2\varphi}}{G_{4\varphi}} \right. \\
& \left.   - \frac{\nabla^2\varphi\, G_{4}G_{3\varphi}}{G_{4\varphi}}
    - 2\nabla^2\varphi\, G_{4\varphi}
    + \frac{2(\nabla\varphi)^{2}G_{4}G_{2\varphi X}}{G_{4\varphi}}
    - \frac{2 \nabla^2\varphi (\nabla\varphi)^{2} G_{4} G_{3\varphi X}}{G_{4\varphi}}
    - \frac{(\nabla\varphi)^{2} G_{4} G_{3\varphi\varphi}}{G_{4\varphi}}\right. \\
& \left.  - 2(\nabla\varphi)^{2} G_{4\varphi\varphi}
    - \frac{2 G_{4} G_{3X} \varphi_{ij}^{2}}{G_{4\varphi}}
    - \frac{4 G_{4} G_{3X} \varphi_{iij}\varphi_{j}}{G_{4\varphi}}
    - 2 G_{3X} \varphi_{i}\varphi_{ij}\varphi_{j}
    + \frac{4 G_{4} G_{2XX} \varphi_{i}\varphi_{ij}\varphi_{j}}{G_{4\varphi}} \right. \\
& \left. - \frac{4 \nabla^2\varphi\, G_{4} G_{3XX}\varphi_{i}\varphi_{ij}\varphi_{j}}{G_{4\varphi}}
    - \frac{4 G_{4} G_{3\varphi X}\varphi_{i}\varphi_{ij}\varphi_{j}}{G_{4\varphi}}
    - \frac{4 G_{4} G_{3XX} \varphi_{ij}\varphi_{ik}\varphi_{j}\varphi_{k}}{G_{4\varphi}}
    + \frac{G_{3} G_{4} \varphi_{j} \varphi_{iij}}{G_{4\varphi}\, \varphi_{i}^{2}} \right. \\
& \left.- \frac{G_{3} G_{4} \varphi_{j} \Psi_{iij}}{G_{4\varphi}\, \varphi_{i}^{2}}
    \right]
    + \frac12\left[
    -\frac{\nabla^2\varphi\, G_{3} G_{4}}{G_{4\varphi}\, \varphi_{i}}
    - \frac{(\nabla\varphi)^{2} G_{4} G_{3\varphi}}{G_{4\varphi}\varphi_{i}}
    + \frac{2 G_{4} G_{2X} \varphi_{i}}{G_{4\varphi}}
    - \frac{4\nabla^2\varphi\, G_{4} G_{3X} \varphi_{i}}{G_{4\varphi}}\right. \\
& \left. - \frac{G_{4} G_{3\varphi} \varphi_{i}}{G_{4\varphi}}
    - \frac{2(\nabla\varphi)^{2}G_{4}G_{3\varphi X} \varphi_{i}}{G_{4\varphi}}
    - \frac{6 G_{4} G_{3X} \varphi_{ij}\varphi_{j}}{G_{4\varphi}}
    - \frac{4 G_{4} G_{3XX} \varphi_{i}\varphi_{j}\varphi_{k}\varphi_{kj}}{G_{4\varphi}}
    \right] \Psi_{i} \\
&   + \frac12\left[
    -\frac{G_{3} G_{4} \varphi_{j}}{G_{4\varphi}\varphi_{i}}
    - \frac{2 G_{4} G_{3X} \varphi_{i}\varphi_{j}}{G_{4\varphi}}
    \right]\Psi_{ij} 
    - \frac12\left[
    \frac{\nabla^2\varphi\, G_{3} G_{4}}{G_{4\varphi}\varphi_{i}}
    + \frac{(\nabla\varphi)^{2} G_{4} G_{3\varphi}}{G_{4\varphi}\varphi_{i}}
    - 2(\nabla\varphi)^{2} G_{3X} \varphi_{i}\right. \\
& \left. - \frac{2 G_{4} G_{2X} \varphi_{i}}{G_{4\varphi}}
    + \frac{4 G_{4} G_{2XX} \varphi_{i}}{G_{4\varphi}}
    - \frac{4\nabla^2\varphi\,(\nabla\varphi)^{2} G_{4} G_{3XX} \varphi_{i}}{G_{4\varphi}}
    + \frac{G_{4} G_{3\varphi} \varphi_{i}}{G_{4\varphi}}
    + 2 G_{4\varphi}\varphi_{i} \right. \\
& \left. - \frac{2(\nabla\varphi)^{2}G_{4}G_{3\varphi X}\varphi_{i}}{G_{4\varphi}}
    - \frac{8 G_{4}G_{3XX}\varphi_{i}\varphi_{ij}}{G_{4\varphi}}
    - \frac{2 G_{4} G_{3X} \varphi_{ij}\varphi_{j}}{G_{4\varphi}}
    + \frac{4 G_{4} G_{3XX}\varphi_{i}\varphi_{j}\varphi_{k}\varphi_{kj}}{G_{4\varphi}}
    \right]\Phi_{i} \\
&   - \frac12\left[
    \frac{G_{3} G_{4} \varphi_{j}}{G_{4\varphi}\varphi_{i}}
    + \frac{2 G_{4} G_{3X} \varphi_{i}\varphi_{j}}{G_{4\varphi}}
    \right]\Phi_{ij}
    + \frac{1}{2}\left[
    G_{2}
    - (\nabla\varphi)^{2} G_{3\varphi}
    + \frac{2\,\nabla^2\varphi\, G_{4}\, G_{2X}}{G_{4\varphi}} \right. \\
& \left. - \frac{2\,(\nabla^2\varphi)^{2}\, G_{4}\, G_{3X}}{G_{4\varphi}}
    - \frac{G_{4}\, G_{2\varphi}}{G_{4\varphi}}
    - \frac{\nabla^2\varphi\, G_{4}\, G_{3\varphi}}{G_{4\varphi}}
    - 2\,\nabla^2\varphi\, G_{4\varphi}
    + \frac{2(\nabla\varphi)^{2} G_{4}\, G_{2\varphi X}}{G_{4\varphi}} \right. \\
& \left. - \frac{2\,\nabla^2\varphi\, (\nabla\varphi)^{2} G_{4}\, G_{3\varphi X}}{G_{4\varphi}}
    - \frac{(\nabla\varphi)^{2} G_{4}\, G_{3\varphi\varphi}}{G_{4\varphi}}
    - 2(\nabla\varphi)^{2} G_{4\varphi\varphi}
    - \frac{2 G_{4}\, G_{3X}\, \varphi_{ij}^{2}}{G_{4\varphi}}
    - \frac{4 G_{4}\, G_{3X}\, \varphi_{iij}\,\varphi_{j}}{G_{4\varphi}} \right. \\
& \left. - 2 G_{3X}\,\varphi_{i}\,\varphi_{ij}\,\varphi_{j} 
    + \frac{4 G_{4}\, G_{2XX}\,\varphi_{i}\,\varphi_{ij}\,\varphi_{j}}{G_{4\varphi}}
    - \frac{4\,\nabla^2\varphi\, G_{4}\, G_{3XX}\,\varphi_{i}\,\varphi_{ij}\,\varphi_{j}}{G_{4\varphi}}
    - \frac{4 G_{4}\, G_{3\varphi X}\,\varphi_{i}\,\varphi_{ij}\,\varphi_{j}}{G_{4\varphi}} \right. \\
& \left. - \frac{4 G_{4}\, G_{3XX}\,\varphi_{ij}\,\varphi_{ik}\,\varphi_{j}\,\varphi_{k}}{G_{4\varphi}}
    + \frac{G_{3}\, G_{4}\, \varphi_{j}\, \Phi_{iij}}{G_{4\varphi}\,\varphi_{i}^{2}}
    - \frac{G_{3}\, G_{4}\, \varphi_{j}\, \Psi_{iij}}{G_{4\varphi}\,\varphi_{i}^{2}}
    \right] = 4\pi G\rho. \numberthis
    \label{weak field limit Horndeski}
\end{align*}
We want this long equation to reduce to the much simpler classical RG equation, i.e.
\begin{equation}
    \nabla\cdot[\epsilon(\varphi,X)\nabla\Psi] = \nabla\epsilon\cdot\nabla\Psi + \epsilon\nabla^2\Psi = \epsilon_\varphi\nabla\varphi\cdot\nabla\Psi + \epsilon_X \nabla X\cdot\nabla\Psi + \epsilon\nabla^2\Psi = 4\pi G\rho,
    \label{classical RG}
\end{equation}
where $\epsilon$ is the permittivity function. We assume a generic dependence of the permittivity on the scalar field and its kinetic term, which we leave implicit. It is convenient to substitute explicitly $\nabla X$ in the previous expression. Since, to first order in the perturbations, $X=(1-2\Phi)(\nabla\varphi)^2$, we get
\begin{equation}
    \partial_iX = -2\partial_i\Phi(\nabla\varphi)^2 + 2(1-2\Psi)\partial_j\partial_i\varphi\partial_j\varphi,
    \label{nabla X}
\end{equation}
and then Eq. (\ref{classical RG}) becomes
\begin{equation}
\epsilon_\varphi\nabla\varphi\cdot\nabla\Psi + 2\epsilon_X\partial_j\partial_i\varphi\partial_j\varphi\partial_i\Psi + \epsilon\nabla^2\Psi = 4\pi G\rho,
    \label{classical RG 2}
\end{equation}
where we discarded second order terms.  
To have Eq. (\ref{weak field limit Horndeski}) reduce to Eq. (\ref{classical RG 2}) it is sufficient for the coefficients of the various terms in the two equations to be equal term by term: for instance, the coefficients multiplying $\nabla\varphi\cdot\nabla\Phi$ -- or $\nabla^2\Phi$, or $\nabla^2\Psi$, or any other derivative operator applied to one of the fields -- in the two equations must coincide. This is why in Eq. (\ref{weak field limit Horndeski}) we grouped terms multiplying the scalar potentials $\Phi$ and $\Psi$ and their first and second derivatives. There is only one caveat to this reasoning: since the explicit expression for the kinetic term $X$ contains the scalar perturbation $\Phi$, the $X$-dependence of the $G$ functions in the last parentheses of Eq. (\ref{weak field limit Horndeski}) -- i.e. the part of the equation which is apparently not proportional to any scalar perturbation -- might give some terms that are actually proportional to $\Phi$. Therefore, when equating the coefficients of $\Phi$ in the two equations we must consider this fact carefully. On the other hand, this caveat has no influence on any other terms -- i.e. any term already multiplying a scalar perturbation or its derivatives -- since any possible term containing $\Phi$ coming from the $G$ functions in this case will be second order and thus negligible.

Eq. (\ref{weak field limit Horndeski}) contains many more terms than its RG counterpart, and we thus have many constraints: the first and most notable one comes from the term multiplying $\nabla^2\Psi$, which yields $\epsilon = G_4$. This fact has a clear physical interpretation: since $G_4$ is the function providing the gravity-scalar coupling in the Horndeski action, the gravitational boost effect, measured by the permittivity in the classical description of RG, is directly linked to this function.  

The second thing to notice is that we do not have any second derivative of $\Psi$ in the RG equation, therefore we can set the coefficients of $\nabla^2\Psi$ and $\Psi_{ij}$ to zero: this implies $G_3 = 0$. This condition implies a huge simplification that erases many terms in Eq. (\ref{weak field limit Horndeski}), which becomes
\begin{equation}
    \begin{aligned}
&   \nabla^2\Psi G_{4}
    + \frac{G_{4} G_{2X} \varphi_{i}}{G_{4\varphi}}\Psi_{i} 
    - \frac12\left[
    - \frac{2 G_{4} G_{2X} \varphi_{i}}{G_{4\varphi}}
    + \frac{4 G_{4} G_{2XX} \varphi_{i}}{G_{4\varphi}}
    + 2 G_{4\varphi}\varphi_{i} 
    \right]\Phi_{i} \\
&   - \Phi\left[
    \frac{2\,\nabla^2\varphi\, G_{4}\, G_{2X}}{G_{4\varphi}} 
    - 2\,\nabla^2\varphi\, G_{4\varphi}
    + \frac{2 (\nabla\varphi)^{2} G_{4} G_{2\varphi X}}{G_{4\varphi}}
    - 2(\nabla\varphi)^{2} G_{4\varphi\varphi} 
    + \frac{8 G_{4} G_{2XX} \varphi_{i}\varphi_{ij}\varphi_{j}}{G_{4\varphi}} 
    \right] \\
&   + \Psi\left[
    G_{2}
    + \frac{2\nabla^2\varphi\, G_{4} G_{2X}}{G_{4\varphi}}
    - \frac{G_{4}G_{2\varphi}}{G_{4\varphi}}
    - 2\nabla^2\varphi\, G_{4\varphi} 
    + \frac{2(\nabla\varphi)^{2}G_{4}G_{2\varphi X}}{G_{4\varphi}} 
    - 2(\nabla\varphi)^{2} G_{4\varphi\varphi} \right. \\
& \left. + \frac{4 G_{4} G_{2XX} \varphi_{i}\varphi_{ij}\varphi_{j}}{G_{4\varphi}}
    \right] 
    + \frac12\left[
    G_{2}
    + \frac{2\nabla^2\varphi\, G_{4} G_{2X}}{G_{4\varphi}}
    - \frac{G_{4}G_{2\varphi}}{G_{4\varphi}}
    - 2\nabla^2\varphi\, G_{4\varphi} 
    + \frac{2(\nabla\varphi)^{2}G_{4}G_{2\varphi X}}{G_{4\varphi}} \right. \\
& \left. - 2(\nabla\varphi)^{2} G_{4\varphi\varphi} 
    + \frac{4 G_{4} G_{2XX} \varphi_{i}\varphi_{ij}\varphi_{j}}{G_{4\varphi}}
    \right] = 4\pi G\rho.
\end{aligned}
\label{weak field limit Horndeski 2}
\end{equation}
Next, equating the coefficients of $\Psi_i$ in Eqs. (\ref{weak field limit Horndeski 2}) and (\ref{classical RG 2}) tells us that $G_4G_{2X}/G_{4\varphi} = G_{4\varphi}$. This relation further simplifies Eq. (\ref{weak field limit Horndeski 2}) to
\begin{equation}
    \begin{aligned}
&   \nabla^2\Phi G_{4}
    + \Psi\left[
    \frac{2 (\nabla\varphi)^{2} G_{4} G_{2\varphi X}}{G_{4\varphi}}
    - 2(\nabla\varphi)^{2} G_{4\varphi\varphi} + \frac{8 G_{4} G_{2XX} \varphi_{i}\varphi_{ij}\varphi_{j}}{G_{4\varphi}} 
    \right] \\
&   + \Phi\left[
    G_{2}
    - \frac{G_{4}G_{2\varphi}}{G_{4\varphi}}
    + \frac{2(\nabla\varphi)^{2}G_{4}G_{2\varphi X}}{G_{4\varphi}} 
    - 2(\nabla\varphi)^{2} G_{4\varphi\varphi} 
    + \frac{4 G_{4} G_{2XX} \varphi_{i}\varphi_{ij}\varphi_{j}}{G_{4\varphi}}
    \right] \\
&   + \frac12\left[
    G_{2}
    - \frac{G_{4}G_{2\varphi}}{G_{4\varphi}}
    + \frac{2(\nabla\varphi)^{2}G_{4}G_{2\varphi X}}{G_{4\varphi}} - 2(\nabla\varphi)^{2} G_{4\varphi\varphi} 
    + \frac{4 G_{4} G_{2XX} \varphi_{i}\varphi_{ij}\varphi_{j}}{G_{4\varphi}}
    \right] \\
&   + \frac{G_{4} G_{2X} \varphi_{i}}{G_{4\varphi}}\Phi_{i} 
    + \frac{2 G_{4} G_{2XX} \varphi_{i}}{G_{4\varphi}}\Psi_{i} = 4\pi G\rho.
\end{aligned}
\label{weak field limit Horndeski 3}
\end{equation}
Now, paying attention to the fact that, to first order, $X\approx(1-2\Phi)(\nabla\varphi)^2 \implies (\nabla\varphi)^2 \approx (1+2\Phi)X$, we recognize that the third and fourth terms in the "zeroth order parentheses" in Eq. (\ref{weak field limit Horndeski 3}) (the last pair of square brackets) give two terms proportional to $\Phi$ that cancel the first two terms inside the parentheses already multiplying $\Phi$. Then, equating both the 
$\Phi$ and $\Phi_i$ coefficients to zero requires $G_{2XX}=0$, leaving us with one final condition:
\begin{equation}
    G_{2}
    - \frac{G_{4}G_{2\varphi}}{G_{4\varphi}}
    + \frac{2G_{4}G_{2\varphi X}}{G_{4\varphi}}X - 2G_{4\varphi\varphi}X = 0.
\end{equation}
Thus, we have obtained the whole set of conditions given in Eqs. (\ref{conditions}).

\section{Background evolution equations in gCRG}
\label{Background evolution equations in gCRG}
In this section, we derive the explicit time evolution for the background value of the scalar field, $\varphi(a)$, and for the Hubble-Lemaître parameter, $H(a)$, in the case of a flat and matter-dominated universe described by the metric in Eq. (\ref{FRW}). In this metric, Eqs. (\ref{Rearrenged Modified Einsten}) and (\ref{Rearrenged KG}) yield the following equations,
\begin{equation}
    \left(\frac{G_{4\varphi}^2}{G_4}-G_{4\varphi\varphi}\right)\dot{\varphi}^2 - G_{4\varphi}\ddot{\varphi} - 3G_4\frac{\ddot{a}}{a} = 4\pi G\rho
\end{equation}
\begin{equation}
    G_{4\varphi}H\dot{\varphi} + G_4\left(2H^2 + \frac{\ddot{a}}{a}\right) = 0
\end{equation}
\begin{equation}
    G_{4\varphi}\left(-\ddot{\varphi} - 3H\dot{\varphi}\right) - 2\Xi G_{4} + G_{4\varphi\varphi}\dot{\varphi}^2 = - 4\pi G \rho
\end{equation}
where $H=\dot{a}/a$, and a dot denotes a time derivative. Introducing the useful parameters 
\begin{equation}
    \Omega \equiv \frac{\Omega_{\mathrm{m}}}{G_4} \equiv \frac{8 \pi G}{3 H^{2} G_4} \rho_{\mathrm{m}}\, , \qquad \Omega_{\Xi}=\frac{\Xi}{3 H^{2}} \,,
    \label{omega, omega_xi}
\end{equation}
\begin{equation}
    q \equiv-\frac{a \ddot{a}}{\dot{a}^{2}}=-\frac{\ddot{a}}{a H^{2}}\, , \qquad q_{\varphi} \equiv-\frac{G_{4\varphi}}{G_4}\frac{\ddot{\varphi}}{H^{2}} \, , \qquad \zeta_\varphi \equiv \frac{G_{4\varphi}}{G_4}\frac{\dot{\varphi}}{H}\,, \qquad \zeta_{\varphi\varphi} \equiv \frac{G_{4\varphi\varphi}}{G_4}\frac{\dot{\varphi}^2}{H \varphi} \, ,
\end{equation}
the system of equations simplifies to
\begin{equation}
    \begin{aligned}
q+\frac{1}{3} \zeta_\varphi^{2} - \frac{1}{3}\zeta_{\varphi\varphi} + \frac{1}{3} q_{\varphi} & =\frac{1}{2} \Omega\,, \\
2+\zeta_\varphi+\frac{1}{3} \zeta_\varphi^{2} - \frac{1}{3}\zeta_{\varphi\varphi} + \frac{1}{3} q_{\varphi} & =\frac{1}{2} \Omega\,, \\
-\frac{1}{3} q_{\varphi}+\zeta_\varphi + \frac{1}{3}\zeta_{\varphi\varphi} &  = \frac{1}{2} \Omega-2 \Omega_{\Xi} \,.
\label{passaggio 1}
\end{aligned}
\end{equation}
Subtracting the first equation from the second in the system (\ref{passaggio 1}),  we find
\begin{equation}
    q - \zeta = 2 \implies \frac{d}{d t} \ln G_4=-\frac{d}{d t}(\ln \dot{a}+2 \ln a) \implies
    G_4=\frac{H_{0} G_{4,0}}{H a^{3}}\,,
    \label{G_4 analitico}
\end{equation}
where $H_0$ and $G_{4,0}$ are the values of the Hubble-Lema\^\i tre parameter and the boost factor at the present time. Subtracting the first equation from the third in Eq. (\ref{passaggio 1}) we get instead $q=\Omega-2 \Omega_{\Xi}-\zeta-\zeta^{2}/3$. Substituting Eq.~(\ref{G_4 analitico}) into this relation and the two definitions of Eqs.~(\ref{omega, omega_xi}) yield
\begin{equation}
    \left(\frac{d}{d t} \ln \dot{a}-\frac{d}{d t} \ln a\right)^{2}=\left(\frac{d}{d t} \ln \frac{\dot{a}}{a}\right)^{2} \implies \frac{d H}{d t}= \pm \sqrt{3}\left(H^{4}+\Omega_{0} H_{0} H^{3}-2 \Omega_{\Xi 0} H_{0}^{2} H^{2}\right)^{1 / 2}\,,
    \label{dH/dt}
\end{equation}
where $\Omega_0$ and $\Omega_{\Xi 0}$ are the cosmological parameters at the present time. The last differential equation can be directly integrated to find 
\begin{equation}
    H(a) = -\frac{1}{2}\Omega_0 H_0 + \sqrt{\frac{1}{4}\Omega_0^2 + 2\Omega_{\Xi 0}}\frac{H_0}{2}\left[ \left(C + \sqrt{C^2 -1}\right)a^{\pm\sqrt{3}} + \left(C + \sqrt{C^2 -1}\right)^{-1}a^{\mp\sqrt{3}} \right]\,,
\end{equation}
where the inequality
\begin{equation}
    C \equiv \frac{1 + \frac{\Omega_0}{2}}{\sqrt{\frac{1}{4}\Omega_0^2 + 2\Omega_{\Xi 0}}} \geq 1\,
    \label{omega_xi condition}
\end{equation}
must be satisfied to guarantee that $H(a)$ is real. 

\section{Linear perturbation theory in gCRG}
\subsection{Growth factor equation}
\label{Linear perturbation theory in gCRG}
In this section, we want to develop the linear perturbation theory for the Horndeski theory we are studying, in order to obtain the growth factor equation for the density contrast in a gCRG universe. First of all, we perturb the Einstein and KG equations, Eqs. (\ref{Rearrenged Modified Einsten}) and (\ref{Rearrenged KG}), obtaining
\begin{align*}
& G_{4\varphi} \delta  \varphi R_{\enspace\nu}^{\mu} + G_4 \delta R_{\enspace\nu}^{\mu} - G_{4\varphi}\left[\delta g^{\mu \lambda} \partial_{\lambda} \partial_{\nu} \varphi + g^{\mu \lambda} \partial_{\lambda} \partial_{\nu} \delta \varphi - \delta g^{\mu \lambda} \Gamma_{\enspace\lambda \nu}^{\rho} \partial_{\rho} \varphi -  \right.\\
& \left. g^{\mu \lambda} \delta \Gamma_{\enspace \lambda \nu}^{\rho} \partial_{\rho} \varphi - g^{\mu \lambda} \Gamma_{\enspace \lambda \nu}^{\rho} \partial_{\rho} \delta \varphi\right] +\left(\frac{2G_{4\varphi}G_{4\varphi\varphi}}{G_4}  - \frac{G_{4\varphi}^3}{G_4^2} - G_{4\varphi\varphi\varphi}\right)\delta \varphi g^{\mu \lambda} \partial_{\lambda} \varphi \partial_{\nu} \varphi \\
& + \left(\frac{G_{4\varphi}^2}{G_4} - G_{4\varphi\varphi}\right) \delta g^{\mu \lambda} \partial_{\lambda} \varphi \partial_{\nu} \varphi + \left(\frac{G_{4\varphi}^2}{G_4} - G_{4\varphi\varphi}\right) g^{\mu \lambda}\left( \partial_{\lambda} \delta \varphi \partial_{\nu} \varphi + \partial_{\lambda} \varphi \partial_{\nu} \delta \varphi \right) = 4 \pi G \delta T_{\enspace \nu}^{\mu}\,, \numberthis
\label{perturbed einstein}
\end{align*}
\begin{align*}
    & G_{4\varphi\varphi}\delta\varphi\square\varphi +  G_{4\varphi}\left[\delta g^{\alpha \beta} \partial_{\alpha} \partial_{\beta} \varphi+g^{\alpha \beta} \partial_{\alpha} \partial_{\beta} \delta \varphi-\delta g^{\alpha \beta} \Gamma_{\enspace \alpha \beta}^{\lambda} \partial_{\lambda} \varphi-g^{\alpha \beta} \delta \Gamma_{\enspace \alpha \beta}^{\lambda} \partial_{\lambda} \varphi-g^{\alpha \beta} \Gamma_{\enspace \alpha \beta}^{\lambda} \partial_{\lambda} \delta \varphi\right] \\
    & \quad -2\Xi G_{4\varphi}\delta\varphi  + G_{4\varphi\varphi\varphi}X\delta\varphi+ 2G_{4\varphi\varphi}\partial^\alpha\varphi\partial_\alpha\delta\varphi= 4\pi G \left( \delta g^{\mu \nu} T_{\mu \nu}+g^{\mu \nu} \delta T_{\mu \nu}\right)\,. \numberthis
    \label{perturbed klein-gordon}
\end{align*}
The perturbed stress-energy tensor that appears in these expressions is 
\begin{equation}
    \delta T_{\mu \nu}=(\delta \rho+\delta P) u_{\mu} u_{\nu}+(\rho+P)\left[\delta u_{\mu} u_{\nu}+u_{\mu} \delta u_{\nu}\right]+\delta P g_{\mu \nu}+P \delta g_{\mu \nu}+\pi_{\mu \nu}\,,
\end{equation}
Finally, we compute the first order components of the perturbed Christoffel symbols and Ricci tensor from the metric in Eq. (\ref{perturbed FLRW metric}) and substitute them in Eq. (\ref{perturbed einstein}), which yields the following four field equations (corresponding, respectively, to the $00$, $0i$, traceless $ij$ and trace $ij$ components of the Einstein equations)
\begin{align*}
    & G_4 \left[6 \mathcal{H}^\prime \Psi-3 \Phi^{\prime\prime}+\nabla^{2} \Psi -3 \mathcal{H}(\Phi^{\prime}-\Psi^\prime)\right] + \left( G_{4\varphi\varphi}\mathcal{H}\varphi^\prime - G_{4\varphi\varphi}\varphi^{\prime\prime} - 3 \mathcal{H}^\prime G_{4\varphi}\right) \delta \varphi \\
    & + G_{4\varphi}\left[ 2 \Psi \varphi^{\prime\prime}-\delta \varphi^{\prime\prime}-2 \mathcal{H} \varphi^\prime \Psi + \varphi^\prime \Psi^\prime + \mathcal{H} \delta \varphi^\prime\right] + \left(\frac{2G_{4\varphi}G_{4\varphi\varphi}}{G_4}  -\frac{G_{4\varphi}^3}{G_4^2} - G_{4\varphi\varphi\varphi}\right)\delta \varphi \left(\varphi^\prime\right)^{2} \\
    & + 2\left(\frac{G_{4\varphi}^2}{G_4} - G_{4\varphi\varphi}\right)\left[ - \Psi\left(\varphi^\prime\right)^{2} + \varphi^\prime \delta \varphi^\prime\right] = 4 \pi G a^{2} \delta \rho\,, \numberthis
    \label{00 perturbed einstein}
\end{align*}
\begin{align*}
    & 2G_4\left(\partial_i\Phi^\prime - \mathcal{H}\partial_i\Psi\right) + G_{4\varphi}\left[\partial_i\delta\varphi^\prime - \partial_i\Psi\varphi^\prime - \mathcal{H}\partial_i\delta\varphi\right] - \left(\frac{G_{4\varphi}^2}{G_4} - G_{4\varphi\varphi}\right)\varphi^\prime\partial_i\delta\varphi \\
    & = 4\pi G a^2\left(\rho + P\right)v_i\,, \numberthis
    \label{0i perturbed einstein}
\end{align*}
\begin{equation}
    G_4\partial^{i} \partial_{j}\left(\Psi+\Phi\right) +G_{4\varphi}\partial^{i} \partial_{j}\delta \varphi = -4 \pi G a^{2} \partial^{i} \partial_{j} \Pi \,;
    \label{ij traceless perturbed einstein}
\end{equation}
\begin{align*}
    & G_{4\varphi}\delta \varphi\left(\mathcal{H}^\prime+2 \mathcal{H}^{2}\right)+ G_4\left(-\nabla^{2} \Phi - 2 \mathcal{H}^\prime \Psi + \Phi^{\prime\prime} - \mathcal{H} \Psi^\prime + 5 \mathcal{H} \Phi^\prime - 4 \mathcal{H}^{2} \Psi\right) - 2 \Phi \mathcal{H} G_{4\varphi}\varphi^\prime \\
    & + \varphi^\prime[2 \mathcal{H}(\Phi-\Psi)+ \Phi^\prime] +\mathcal{H} \delta \varphi^\prime - \frac{G_4}{3} \nabla^{2}\left(\Psi+\Phi\right)  - G_{4\varphi}\frac{\nabla^{2}\delta \varphi}{3\varphi}= 4\pi G a^{2} \delta P \, . \numberthis
    \label{ij trace perturbed einstein}
\end{align*}
In these equations, $\mathcal{H}=a^\prime/a$ and a prime indicates differentiation with respect to the conformal time. $\Pi$ is the scalar function that characterizes the spatial part of the anisotropic stress tensor, i.e., $\pi_{ij} = a^2\left(\partial_i\partial_j\Pi - \delta_{ij}\nabla^2\Pi/3\right)$. The KG equation in Eq. (\ref{perturbed klein-gordon}) turns out to be 
\begin{equation}
    \begin{aligned}
        & G_{4\varphi\varphi}\delta\varphi \left(-\varphi^{\prime\prime} - 2\mathcal{H}\varphi^\prime\right) + G_{4\varphi} \left[ 2 \Psi \varphi^{\prime\prime}-\delta \varphi^{\prime\prime}+\nabla^{2} \delta \varphi + 4 \mathcal{H} \varphi^\prime \Psi + \varphi^\prime ( \Psi^\prime-3 \Phi^\prime)-2 \mathcal{H} \delta\varphi^\prime \right]\\
        & - 2 \Xi G_{4\varphi} a^{2} \delta \varphi - G_{4\varphi\varphi\varphi}\left(\varphi^\prime\right)^2\delta\varphi - 2G_{4\varphi\varphi}\varphi^\prime\delta\varphi^\prime + 2G_{4\varphi\varphi}\Psi(\varphi^\prime)^2 = 4 \pi G a^{2}(3 \delta P-\delta \rho) \,.
    \label{perturbed KG 2}
    \end{aligned}
\end{equation}

Finally, perturbing the conservation equation, $\nabla_\mu T^\mu_{\enspace\nu} = 0$, gives the two equations
\begin{equation}
    \delta^\prime + 3 \mathcal{H}(c_s^2 - w)\delta +(1+w)\left(\theta + 3\Phi^\prime\right)=0 \,,
    \label{perturbed system continuity}
\end{equation}
\begin{equation}
    \rho(1+w)\left( \mathcal{H} \theta + \theta^\prime + \nabla^2 \Psi -3w\mathcal{H}\theta\right) + c_s^2\rho\nabla^2\delta + \frac{2}{3}\nabla^4\Pi = 0\,,
    \label{perturbed system Euler}
\end{equation}
where we have defined the equation of state, $P = w\rho$, the speed of sound $c_s^2 = \delta P/\delta\rho$, the density contrast $\delta = \delta\rho/\rho$ and the velocity divergence $\theta = \partial_i v^i$. 

In order to determine whether this theory can describe the formation of cosmic structure, we must derive an equation for the evolution of the density contrast, $\delta$. Since structure formation takes place at relatively late times, well into matter domination, we can set $w=c_s^2=0$. Eqs. (\ref{perturbed system continuity}) and (\ref{perturbed system Euler}) thus become
\begin{align}
    \delta^\prime = & - \left(\theta + 3\Phi^\prime\right)\,, \label{continuità}\\ 
    \theta^\prime = & - \mathcal{H}\theta - \nabla^2\Psi\,.   \label{eulero}
\end{align}
The first step is to combine these two equations: we switch to a derivative with respect to the variable $N=\ln a$, we derive the first equation and substitute the second equation into the first one; we find
\begin{equation}
    \delta^{\prime\prime} + \left(2+\frac{H^\prime}{H}\right)\delta^\prime = -6\Phi^\prime + \frac{\nabla^2\Psi}{(aH)^2} - 3\frac{H^\prime}{H}\Phi^\prime - 3\Phi^{\prime\prime}.
    \label{passaggio growth factor}
\end{equation}
where now $^\prime={\rm d}/{\rm d}N$. Next, we substitute Eq. (\ref{perturbed KG 2}) into Eq. (\ref{00 perturbed einstein}) to eliminate $\varphi^{\prime\prime}$, and then we substitute the resulting equation into Eq. (\ref{passaggio growth factor}) to eliminate $\nabla^2\Psi$. The result is the growth factor equation, Eq. (\ref{delta evolution}). By performing the same change of variable in the perturbed KG equation, Eq. (\ref{perturbed KG 2}), and in the background KG equation, Eq. (\ref{Rearrenged KG}), we obtain Eqs. (\ref{perturbed KG}) and (\ref{background KG}). 

\subsection{Numerical solution}
\label{numerical solution}
In this section, we show how to approach the numerical solution of the set of equations given by Eqs. (\ref{delta evolution}), (\ref{perturbed KG}) and (\ref{background KG}), which we report here under the assumptions of the QSA, meaning that we neglect the terms containing the scalar potentials $\Phi$ and $\Psi$ and their derivatives: 
\begin{subequations} \label{differential system}
\begin{align*}
    & \delta^{\prime\prime} + \left(2 + \frac{H^\prime}{H}\right)\delta^\prime - \frac{8\pi G}{G_4}\frac{\rho_{m0}}{H^2a^3}\delta - \frac{\nabla^2\delta\varphi}{a^2H^2}\frac{G_{4\varphi}}{G_4} + 2\frac{\Xi}{H^2}\delta\varphi\frac{G_{4\varphi}}{G_4} +   3\frac{G_{4\varphi}}{G_4}\left(1+\frac{H^\prime}{H}\right)\delta\varphi \\
    = \; & - 3\frac{G_{4\varphi}}{G_4}\delta\varphi^\prime - 3\frac{G_{4\varphi\varphi}}{G_4}\varphi^\prime\delta\varphi - \left(\frac{2G_{4\varphi}G_{4\varphi\varphi}}{G_4^2} - \frac{G_{4\varphi}^3}{G_4^3}\right)\left(\varphi^\prime\right)^2\delta\varphi - 2\frac{G_{4\varphi}^2}{G_4^2}\varphi^\prime\delta\varphi^\prime  \, ; \numberthis
    \label{equation for delta evolution}
\end{align*}
\begin{align*}
    & \delta \varphi^{\prime\prime} + \left(3+\frac{H^\prime}{H}\right)\delta\varphi^\prime - \frac{4\pi G}{G_{4\varphi}}\frac{\rho_{m0}}{H^2a^3}\delta - \frac{\nabla^{2} \delta\varphi}{a^2H^2} + 2\frac{\Xi}{H^2}\delta\varphi \\
    = \; & - \frac{G_{4\varphi\varphi}}{G_{4\varphi}}\delta\varphi\left[\varphi^{\prime\prime}+\left(3+\frac{H^\prime}{H}\right)\varphi^\prime\right] - \frac{G_{4\varphi\varphi\varphi}}{G_{4\varphi}}\left(\varphi^\prime\right)^2\delta\varphi - 2\frac{G_{4\varphi\varphi}}{G_{4\varphi}}\varphi^\prime\delta\varphi^\prime\,; \numberthis
    \label{equation for delta phi evolution}
\end{align*}
\begin{align*}
     \varphi^{\prime\prime} + \left(3 + \frac{H^\prime}{H}\right)\varphi^\prime - \frac{4\pi G}{G_{4\varphi}}\frac{\rho_{m0}}{H^2a^3} + 2\frac{\Xi}{H^2}\frac{G_4}{G_{4\varphi}} = \frac{G_{4\varphi\varphi}}{G_{4\varphi}}\left(\varphi^\prime\right)^2\,. \numberthis
\end{align*}
\end{subequations}
These three equations yield a coupled system for the three functions $\delta, \delta\varphi$ and $\varphi$. The system also contains other functions of the independent variable $N=\ln a$, in particular the Hubble-Lemaître parameter $H$, the boost factor function $G_4$ and its derivatives with respect to the scalar field. However, in Section \ref{Background evolution equations in gCRG} we have determined analytical solutions for these functions, which are 
\begin{equation}
    H(a) = -\frac{1}{2}\Omega_0 H_0 + \sqrt{\frac{1}{4}\Omega_0^2 + 2\Omega_{\Xi 0}}\frac{H_0}{2}\left[ \left(C + \sqrt{C^2 -1}\right)a^{\pm\sqrt{3}} + \left(C + \sqrt{C^2 -1}\right)^{-1}a^{\mp\sqrt{3}} \right]\,,
\end{equation}
\begin{equation}
     G_4[\varphi(a)]=\frac{H_{0} G_{4,0}}{H(a) a^{3}}.
\end{equation}
Since we have the explicit evolution of $G_4$ with respect to time, we rewrite its derivatives with respect to the scalar field in terms of its time derivatives, which we can compute straight-away:
\begin{equation}
G_{4\varphi} = \frac{G_4^\prime}{\varphi^\prime} \quad \text{where} \quad G_4^\prime = -\frac{G_{4,0}}{H}\,a^{-3}\left(\frac{H^\prime}{H} + 3\right)\,;
\end{equation}
\begin{align*}
    G_{4\varphi\varphi} & = \frac{G_4^{\prime\prime}}{\varphi^{\prime 2}} - \frac{G_4^\prime\,\varphi^{\prime\prime}}{\varphi^{\prime 3}} \quad \text{where} \\
     G_4^{\prime\prime} & = \frac{G_{4,0}}{H}\,a^{-3}\left(\frac{H^\prime}{H}+3\right)^2 - \frac{G_{4,0}}{H}\,a^{-3}\left[\frac{H^{\prime\prime}}{H} - \left(\frac{H^\prime}{H}\right)^2\right]\,; \numberthis
\end{align*}
\begin{align*}
    G_{4\varphi\varphi\varphi} &= \frac{G_4^{\prime\prime\prime}}{\varphi^{\prime 3}} - \frac{2G_4^{\prime\prime}\,\varphi^{\prime\prime}}{\varphi^{\prime 4}} - \frac{G_4^\prime\,\varphi^{\prime\prime\prime}}{\varphi^{\prime 4}} + \frac{3\varphi^{\prime\prime 2}\,G_4^\prime}{\varphi^{\prime 5}} - \frac{\varphi^{\prime\prime}\,G_4^{\prime\prime}}{\varphi^{\prime 4}} \quad \text{where} \\
    G_4^{\prime\prime\prime} &= -\frac{G_{4,0}}{H}\,a^{-3}\left(\frac{H^\prime}{H}+3\right)^3 + 3\frac{G_{4,0}}{H}\,a^{-3}\left(\frac{H^\prime}{H}+3\right)\left[\frac{H^{\prime\prime}}{H}-\left(\frac{H^\prime}{H}\right)^2\right] \notag\\
    & \quad - \frac{G_{4,0}}{H}\,a^{-3}\left[\frac{H^{\prime\prime\prime}}{H} - 3\frac{H^{\prime\prime}H^\prime}{H^2} - 2\left(\frac{H^\prime}{H}\right)^3\right]. \numberthis
\end{align*}
The third derivation of $\varphi$ is found by deriving the background KG equation:
\begin{align*}
    \varphi^{\prime\prime\prime} &= -\frac{9}{4}\frac{\Omega_m\,a^{-3}}{H^2}\frac{\varphi^\prime}{G_4^\prime} + \frac{3}{4}\frac{\Omega_m\,a^{-3}}{H^2}\frac{\varphi^{\prime\prime}}{G_4^\prime} - \frac{3\Omega_X}{H^2}\varphi^\prime - \frac{3\Omega_X}{H^2}\frac{G_4}{G_4^\prime}\varphi^{\prime\prime} - \frac{1}{2}\left(\frac{H^\prime}{H}\right)^2\varphi^\prime \notag\\
&\quad - \frac{1}{2}\frac{H^{\prime\prime}}{H}\varphi^\prime + \frac{1}{2}\frac{H^\prime}{H}\frac{G_4^{\prime\prime}}{G_4^\prime}\varphi^\prime - \frac{5}{2}\frac{H^\prime}{H}\varphi^{\prime\prime} - 3\frac{H^\prime}{H}\varphi^\prime - \frac{3}{2}\frac{G_4^{\prime\prime}}{G_4^\prime}\varphi^\prime - \frac{3}{2}\varphi^{\prime\prime} + \frac{1}{2}\frac{G_4^{\prime\prime\prime}}{G_4^\prime}\varphi^\prime - \frac{1}{2}\frac{G_4^{\prime\prime}}{G_4^\prime}\varphi^{\prime\prime} \,,\\[10pt]
\end{align*}
In order to compute the time derivatives of $G_4$, it sufficient to know the time derivatives of $H$, which are
\begin{align}
    H^\prime &= \sqrt{\frac{1}{4}\Omega_0^2 + 2\Omega_{\Xi 0}}\frac{H_0}{2}\left[\pm\sqrt{3}\left(C+\sqrt{C^2-1}\right)\,a^{\pm\sqrt{3}} \mp \frac{\sqrt{3}\,a^{\mp\sqrt{3}}}{C+\sqrt{C^2-1}}\right]\,, \\
    H^{\prime\prime} & = \sqrt{\frac{1}{4}\Omega_0^2 + 2\Omega_{\Xi 0}}\frac{H_0}{2}\left[3\left(C+\sqrt{C^2-1}\right)\,a^{\pm\sqrt{3}} + \frac{3\,a^{\mp\sqrt{3}}}{C+\sqrt{C^2-1}}\right]\,, \\
    H^{\prime\prime\prime} &= 3H^\prime\,.
\end{align}
Substituting all of these quantities into the differential system in Eqs. (\ref{differential system}), we obtain a set of equations in which the only unknowns are $\delta, \delta\varphi$ and $\varphi$. We can solve this system through a Runge-Kutta Cash-Karp algorithm with an adaptive step size, which calculates fourth- and fifth-order accurate solutions and estimates the error as the difference between the two solutions. However, as already discussed in Section \ref{Linear perturbation theory in gCRG}, this direct method is not really viable because there is no physically-motivated way to impose initial conditions on the values of the scalar field $\varphi$ and its derivative $\varphi^\prime$. This situation contrasts with our previous work \cite{Gervani:2026zqm}, where we just imposed initial conditions using the analytical solution available for $\varphi(a)$. However, in the more general context of gCRG, such a solution is no longer known, as was explained in Section \ref{lin pert th}. Therefore, one may resort to the simplified approach offered by QSA, as was done in Section \ref{QSA numerical solution}.

\section{Horndeski curved universe}
\subsection{Linear perturbation theory in a Horndeski curved universe}
\label{Linear perturbation theory in a Horndeski curved universe}
In this section we derive the Einstein and KG equations for a Horndeski curved universe. We start from the metric in Eq. (\ref{curved perturbed metric}). The resulting Christoffel symbols are, to first order in the perturbations,
\begin{align*}
    \Gamma^{0}_{\,\,\,00} & = \mathcal{H} + \Psi'; \\
    \Gamma^{0}_{\,\,\,0i} & = \partial_i \Psi; \\
    \Gamma^{i}_{\,\,\,00} & = \frac{\partial^i \Psi}{\gamma^2}; \\
    \Gamma^{i}_{\,\,\,j0} & = (\mathcal{H} + \Phi')\,\delta^{i}_{j}; \\
    \Gamma^{0}_{\,\,\,ij} & = \gamma^2\!\left(\mathcal{H} + \Phi' + 2\mathcal{H}(\Phi - \Psi)\right)\delta_{ij}; \\
    \Gamma^{i}_{\,\,\,jk} & = -\frac{\mathcal{K}\gamma}{2}\!\left(x_k\,\delta^{i}_{j} + x_j\,\delta^{i}_{k} - x^i\,\delta_{jk}\right) + \partial_k\Phi\,\delta^{i}_{j} + \partial_j\Phi\,\delta^{i}_{k} - \partial^i\Phi\,\delta_{jk}; \numberthis
\end{align*}
while the Ricci tensor components are
\begin{align*}
    R^0_{\,\,\,0} & = a^{-2}\!\left(3\mathcal{H}' - \frac{\Delta\Psi}{\gamma^2} + \frac{\mathcal{K}}{2\gamma}\,\partial_i\Psi\, x^i + 3\Phi'' - 6\mathcal{H}'\Psi + 3\mathcal{H}(\Phi' - \Psi')\right); \\
    R^0_{\,\,\,i} & = a^{-2}\!\left(2\partial_i\Phi' - 2\mathcal{H}\,\partial_i\Psi\right); \\
    R^i_{\,\,\,0} & = -a^{-2}\!\frac{1}{\gamma^2}\left(2\partial_i\Phi' - 2\mathcal{H}\,\partial_i\Psi\right); \\
    R^i_{\,\,\,j} & = a^{-2}\!\left\{\left[2\mathcal{K} + 2\mathcal{H}^2 + \mathcal{H}' + \frac{8\mathcal{K}\Phi}{\gamma} - 12\mathcal{K}\Phi - \frac{\Delta\Phi}{\gamma^2} - 2\mathcal{K}^2 r^2\Phi + \right. \right. \\
    & \left. \left. \frac{\mathcal{K}}{\gamma}\!\left(\partial_k\Phi\, x^k + \frac{1}{2}\partial_k\Psi\, x^k\right) + \Phi'' - 4\mathcal{H}^2\Psi + 5\mathcal{H}\Phi' - \mathcal{H}\Psi' - 2\Psi\mathcal{H}'\right]\delta^{i}{}_{j} \right. \\
    & \left. - \frac{\partial^i\partial_j\Phi}{\gamma^2} - \frac{\partial^i\partial_j\Psi}{\gamma^2} + \frac{\mathcal{K}}{2\gamma}\!\left(\partial^i\Phi + \partial^i\Psi\right)x_j + \frac{\mathcal{K}\gamma}{2}\!\left(\partial_j\Phi + \partial_j\Psi\right)x^i\right\}, \numberthis
\end{align*}
where $\gamma = (1+\mathcal{K}r^2/4)^{-1}$, $\mathcal{H}=a^\prime/a$, and a prime means derivation with respect to  conformal time. These results are approximated to first order in $\Phi$ and $\Psi$. Using them into the perturbed Einstein and KG equations, Eqs. (\ref{perturbed einstein}) and (\ref{perturbed klein-gordon}), we obtain the following four independent equations. The first three are, respectively, the $00$, $0i$ and $ij$ components of the Einsten equations, which are
\begin{align}
 \label{curved perturbed ein 00}
    &- \frac{3 G_{4\varphi} \mathcal{H} \delta\varphi'}{a^2}
    - \frac{G_{4\varphi}^2 \delta\varphi' \varphi'}{a^2 G_4}
    - \frac{6 G_4 \mathcal{H} \Phi'}{a^2}
    - \frac{3 G_{4\varphi} \varphi' \Phi'}{a^2}
    + \frac{G_{4\varphi} \Delta\delta\varphi}{a^2 \gamma^2}
    - \frac{3 G_{4\varphi} \mathcal{H}^2 \delta\varphi}{a^2} \nonumber\\
    &- \frac{3 G_{4\varphi\varphi} \mathcal{H} \varphi' \delta\varphi}{a^2}
    + \frac{G_{4\varphi}^3 {\varphi'}^2 \delta\varphi}{2 a^2 G_4^2}
    - \frac{G_{4\varphi} G_{4\varphi\varphi} {\varphi'}^2 \delta\varphi}{a^2 G_4}
    + \frac{2 G_4 \Delta\Phi}{a^2 \gamma^2}
    - \frac{7 G_4 \mathcal{K} \,\partial_k\Phi\, x^k}{4 a^2 \gamma}
    - \frac{3 G_4 \mathcal{K} \,\partial_k\Psi\, x^k}{4 a^2 \gamma} \nonumber\\
    &- \frac{G_{4\varphi} \mathcal{K} \,\partial_k\delta\varphi\, x^k}{2 a^2 \gamma}
    - \frac{G_4 \gamma\mathcal{K} \,\partial_k\Phi\, x^k}{4 a^2}
    - \frac{G_4 \gamma\mathcal{K} \,\partial_k\Psi\, x^k}{4 a^2}
    - \frac{3 G_{4\varphi} \mathcal{K} \delta\varphi}{a^2}
    - G_{4\varphi} \delta\varphi \Xi
    + \frac{18 G_4 \mathcal{K} \Phi}{a^2}\nonumber\\
    &- \frac{12 G_4 \mathcal{K} \Phi}{a^2 \gamma} 
    + \frac{3 G_4 r^2 \mathcal{K}^2 \Phi}{a^2}
    + \frac{6 G_4 \mathcal{H}^2 \Psi}{a^2}
    + \frac{6 G_{4\varphi} \mathcal{H} \varphi' \Psi}{a^2}
    + \frac{G_{4\varphi}^2 {\varphi'}^2 \Psi}{a^2 G_4} = 0 \\[12pt]
    \label{curved perturbed ein 0i}
    & \frac{G_{4\varphi} \partial_i\delta\varphi'}{a^2}
    + \frac{2 G_4 \partial_i\Phi'}{a^2}
    - \frac{G_{4\varphi} \mathcal{H} \partial_i\delta\varphi}{a^2}
    - \frac{G_{4\varphi}^2 \varphi' \partial_i\delta\varphi}{a^2 G_4}
    + \frac{G_{4\varphi\varphi} \varphi' \partial_i\delta\varphi}{a^2}
    - \frac{2 G_4 \mathcal{H} \partial_i\Psi}{a^2}
    - \frac{G_{4\varphi} \varphi' \partial_i\Psi}{a^2} = 0 \\[12pt]
    \label{curved perturbed ein ij}
    &-\frac{G_{4\varphi} \nabla^i\partial_j\delta\varphi}{a^2 \gamma^2}
    - \frac{G_4 \nabla^i\partial_j\Phi}{a^2 \gamma^2}
    - \frac{G_4 \nabla^i\partial_j\Psi}{a^2 \gamma^2}
    - \frac{G_{4\varphi} \delta\varphi''}{a^2} \delta^i{}_j
    - \frac{2 G_4 \Phi''}{a^2} \delta^i{}_j
    - \frac{G_{4\varphi} \mathcal{H} \delta\varphi'}{a^2} \delta^i{}_j \nonumber\\
    &+ \frac{G_{4\varphi}^2 \delta\varphi' \varphi'}{a^2 G_4} \delta^i{}_j
    - \frac{2 G_{4\varphi\varphi} \delta\varphi' \varphi'}{a^2} \delta^i{}_j
    - \frac{4 G_4 \mathcal{H} \Phi'}{a^2} \delta^i{}_j
    - \frac{2 G_{4\varphi} \varphi' \Phi'}{a^2} \delta^i{}_j
    + \frac{2 G_4 \mathcal{H} \Psi'}{a^2} \delta^i{}_j
    + \frac{G_{4\varphi} \varphi' \Psi'}{a^2} \delta^i{}_j \nonumber\\
    &+ \frac{G_{4\varphi} \Delta\delta\varphi}{a^2 \gamma^2} \delta^i{}_j
    - \frac{G_{4\varphi} \mathcal{H}^2 \delta\varphi}{a^2} \delta^i{}_j
    - \frac{G_{4\varphi\varphi} \varphi'' \delta\varphi}{a^2} \delta^i{}_j
    - \frac{2 G_{4\varphi} \mathcal{H}' \delta\varphi}{a^2} \delta^i{}_j
    - \frac{G_{4\varphi\varphi} \mathcal{H} \varphi' \delta\varphi}{a^2} \delta^i{}_j \nonumber\\
    &- \frac{G_{4\varphi}^3 {\varphi'}^2 \delta\varphi}{2 a^2 G_4^2} \delta^i{}_j 
    + \frac{G_{4\varphi} G_{4\varphi\varphi} {\varphi'}^2 \delta\varphi}{a^2 G_4} \delta^i{}_j
    - \frac{G_{4\varphi\varphi\varphi} {\varphi'}^2 \delta\varphi}{a^2} \delta^i{}_j
    + \frac{G_4 \Delta\Phi}{a^2 \gamma^2} \delta^i{}_j
    + \frac{G_4 \Delta\Psi}{a^2 \gamma^2} \delta^i{}_j \nonumber\\
    &- \frac{3 G_4 \mathcal{K} \,\partial_k\Phi\, x^k}{4 a^2 \gamma} \delta^i{}_j
    - \frac{3 G_4 \mathcal{K} \,\partial_k\Psi\, x^k}{4 a^2 \gamma} \delta^i{}_j 
    - \frac{G_4 \gamma\mathcal{K} \,\partial_k\Phi\, x^k}{4 a^2} \delta^i{}_j
    - \frac{G_4 \gamma\mathcal{K} \,\partial_k\Psi\, x^k}{4 a^2} \delta^i{}_j
    - \frac{G_{4\varphi} \mathcal{K} \delta\varphi}{a^2} \delta^i{}_j \nonumber\\
    &- G_{4\varphi} \delta\varphi \Xi \delta^i{}_j
    + \frac{6 G_4 \mathcal{K} \Phi}{a^2} \delta^i{}_j
    - \frac{4 G_4 \mathcal{K} \Phi}{a^2 \gamma} \delta^i{}_j 
    + \frac{G_4 r^2 \mathcal{K}^2 \Phi}{a^2} \delta^i{}_j
    + \frac{2 G_4 \mathcal{H}^2 \Psi}{a^2} \delta^i{}_j
    + \frac{2 G_{4\varphi} \varphi'' \Psi}{a^2} \delta^i{}_j \nonumber\\
    &+ \frac{4 G_4 \mathcal{H}' \Psi}{a^2} \delta^i{}_j
    + \frac{2 G_{4\varphi} \mathcal{H} \varphi' \Psi}{a^2} \delta^i{}_j
    - \frac{G_{4\varphi}^2 {\varphi'}^2 \Psi}{a^2 G_4} \delta^i{}_j 
    + \frac{2 G_{4\varphi\varphi} {\varphi'}^2 \Psi}{a^2} \delta^i{}_j
    - \frac{G_{4\varphi} \mathcal{K} \,\partial_k\delta\varphi\, x^k}{2 a^2 \gamma} \partial_j\delta\varphi\, x^i \nonumber\\
    &+ \frac{G_4 \gamma\mathcal{K} \,\partial_k\Phi\, x^k}{2 a^2} \partial_j\Phi\, x^i
    + \frac{G_4 \gamma\mathcal{K} \,\partial_k\Psi\, x^k}{2 a^2} \partial_j\Psi\, x^i
    - \frac{G_{4\varphi} \mathcal{K} \,\partial_k\delta\varphi\, x^k}{2 a^2 \gamma} \partial_i\delta\varphi\, x_j 
    + \frac{G_4 \mathcal{K} \,\partial_k\Phi\, x^k}{2 a^2 \gamma} \partial_i\Phi\, x_j \nonumber\\
    &+ \frac{G_4 \mathcal{K} \,\partial_k\Psi\, x^k}{2 a^2 \gamma} \partial_i\Psi\, x_j = 0.
\end{align}
The fourth and final equation is the perturbed KG equation,
\begin{align}
    &-\frac{2 G_{4\varphi} \delta\varphi''}{a^2}
    - \frac{6 G_4 \Phi''}{a^2}
    - \frac{4 G_{4\varphi} \mathcal{H} \delta\varphi'}{a^2}
    + \frac{2 G_{4\varphi}^2 \delta\varphi' \varphi'}{a^2 G_4}
    - \frac{4 G_{4\varphi\varphi} \delta\varphi' \varphi'}{a^2}
    - \frac{18 G_4 \mathcal{H} \Phi'}{a^2}
    - \frac{6 G_{4\varphi} \varphi' \Phi'}{a^2} \nonumber\\
    &+ \frac{6 G_4 \mathcal{H} \Psi'}{a^2}
    + \frac{2 G_{4\varphi} \varphi' \Psi'}{a^2}
    + \frac{2 G_{4\varphi} \Delta\delta\varphi}{a^2 \gamma^2}
    - \frac{6 G_{4\varphi} \mathcal{H}^2 \delta\varphi}{a^2}
    - \frac{2 G_{4\varphi\varphi} \varphi'' \delta\varphi}{a^2}
    - \frac{6 G_{4\varphi} \mathcal{H}' \delta\varphi}{a^2} \nonumber\\
    & - \frac{4 G_{4\varphi\varphi} \mathcal{H} \varphi' \delta\varphi}{a^2} 
    - \frac{G_{4\varphi}^3 {\varphi'}^2 \delta\varphi}{a^2 G_4^2}
    + \frac{2 G_{4\varphi} G_{4\varphi\varphi} {\varphi'}^2 \delta\varphi}{a^2 G_4}
    - \frac{2 G_{4\varphi\varphi\varphi} {\varphi'}^2 \delta\varphi}{a^2}
    + \frac{4 G_4 \Delta\Phi}{a^2 \gamma^2}
    + \frac{2 G_4 \Delta\Psi}{a^2 \gamma^2} \nonumber\\
    & - \frac{7 G_4 \mathcal{K} \,\partial_k\Phi\, x^k}{2 a^2 \gamma}
    - \frac{5 G_4 \mathcal{K} \,\partial_k\Psi\, x^k}{2 a^2 \gamma} 
    - \frac{G_{4\varphi} \mathcal{K} \,\partial_k\delta\varphi\, x^k}{a^2 \gamma}
    - \frac{G_4 \gamma\mathcal{K} \,\partial_k\Phi\, x^k}{2 a^2}
    - \frac{G_4 \gamma\mathcal{K} \,\partial_k\Psi\, x^k}{2 a^2}
    - \frac{6 G_{4\varphi} \mathcal{K} \delta\varphi}{a^2} \nonumber\\
    &- 2 G_{4\varphi} \delta\varphi \Xi
    + \frac{36 G_4 \mathcal{K} \Phi}{a^2}
    - \frac{24 G_4 \mathcal{K} \Phi}{a^2 \gamma} 
    + \frac{6 G_4 r^2 \mathcal{K}^2 \Phi}{a^2}
    + \frac{12 G_4 \mathcal{H}^2 \Psi}{a^2}
    + \frac{4 G_{4\varphi} \varphi'' \Psi}{a^2}
    + \frac{12 G_4 \mathcal{H}' \Psi}{a^2} \nonumber\\
    &+ \frac{8 G_{4\varphi} \mathcal{H} \varphi' \Psi}{a^2}
    - \frac{2 G_{4\varphi}^2 {\varphi'}^2 \Psi}{a^2 G_4}
    + \frac{4 G_{4\varphi\varphi} {\varphi'}^2 \Psi}{a^2} = 0.
    \label{curved perturbed KG}
\end{align}
When we consider the flat universe limit, i.e. $\mathcal{K}\xrightarrow{}0, \gamma\xrightarrow{}1$, this set of four equations reduces to a set that is equivalent to Eqs. (\ref{00 perturbed einstein})-(\ref{perturbed KG 2}), namely the ones previously considered in Section \ref{lin pert th} and Appendix \ref{Linear perturbation theory in gCRG}. 

The next step is to derive the growth factor equation for the case of a curved universe, i.e one with the metric in Eq. (\ref{curved perturbed metric}). This derivation is easily performed by perturbing the conservation equation, $\nabla_\mu T^{\mu}_{\,\nu}=0$, which yields the continuity equation
\begin{equation}
    \delta^\prime + 3\mathcal{H}(c_s^2-w)\delta + (\theta+3\Phi^\prime)(1+w) - \frac{3\mathcal{K}}{2\gamma}x_j v^j(1+w) = 0,
\end{equation}
and the Euler equation
\begin{equation}
    \rho(1+w)(\mathcal{H}\theta + \theta^\prime + \nabla^2\Psi - 3w\mathcal{H}\theta) + c_s^2\rho\nabla^2\delta + \frac{2}{3}\nabla^2\nabla^2\Pi - \mathcal{K}\gamma x^i\partial_j\nabla^2\Pi -\frac{3}{4}\mathcal{K}^2\gamma^2x_ix^j\pi^i_j = 0,
\end{equation}
which, in the case $c_s^2=w=\Pi=0$, reduce to
\begin{align}
    \delta^\prime & + \theta + 3\Phi^\prime - \frac{3\mathcal{K}}{2\gamma}x_j v^j(1+w) = 0 \\
    \theta^\prime & + \mathcal{H}\theta + \nabla^2\Psi = 0 \, .
\end{align}
We combine these two equations into
\begin{equation}
    \delta^{\prime\prime} + \left(2+\frac{H^\prime}{H}\right)\delta^\prime = -3\Phi^{\prime\prime}-3\left(2+\frac{H^\prime}{H}\right)\Phi^\prime + \frac{\nabla^2\Psi}{\gamma^2(aH)^2} - \frac{\mathcal{K}}{2\gamma}\frac{x^j\partial_j\Psi}{(aH)^2},
\end{equation}
which is the curved extension of Eq. (\ref{passaggio growth factor}). Now we proceed in an identical fashion as in Appendix \ref{Linear perturbation theory in gCRG}: we solve for $\varphi^{\prime\prime}$ using Eq. (\ref{curved perturbed KG}) and for $\nabla^2\Psi$ using Eq. (\ref{curved perturbed ein 00}). The final result is the growth factor equation in a curved universe, Eq. (\ref{delta curved evolution}).

\subsection{QSA in a Horndeski curved universe}
\label{QSA in a Horndeski curved universe}
In this appendix we illustrate the detailed computations related to the application of the QSA to the curved case, as discussed at the end of Section \ref{Horndeski theory in a curved universe}. In particular, following \cite{Gambino2024}, we use a set of perturbed equations which includes the field equations and the evolution equation for the scalar field. We then apply the QSA to this set, to show that the presence of curvature does not influence the final result. 

The curved metric in the Newtonian gauge is
\begin{equation*}
     \mathrm{d}s^2 = -(1+2\Phi)\mathrm{d}t^2 + a^{2}(1-2\Psi)\gamma_{ij} \mathrm{d}x^{i}\mathrm{d}x^{j}\,,
\end{equation*}
with $\gamma_{ij}=\delta_{ij}/(1+\mathcal{K}r^2/4)^2$. Note the different definition of the gravitational potentials $\Phi$ and $\Psi$: here we assume the convention of \cite{Gambino2024}. From this metric, the covariant derivatives are:
\begin{align*}
 D_i D_j=&\, \partial_{i}\partial_{j} -  \frac{\mathcal{K}}{2\gamma}\left[ x_i\partial_j + x_j \partial_i - \delta_{ij} x^l \partial_l \right]\,,\\
 D^2 \Psi = &\, \gamma^2 \nabla^2 \Psi - \frac{\mathcal{K}\gamma}{2}x^i \partial_i \Psi \,, \\
 D^2 \Phi = &\, \gamma^2  \nabla^2 \Phi - \frac{\mathcal{K}\gamma}{2}x^i \partial_i \Phi \,, \\
 D^2 \pi = &\, \gamma^2  \nabla^2 \pi - \frac{\mathcal{K}\gamma}{2}x^i \partial_i \pi \,,
\end{align*}
where $\gamma=\left(1+\mathcal{K}r^2/4\right)$ and $\gamma_{ij} = \delta_{ij}/\gamma^2$, and where $\pi = \delta\varphi/\dot{\varphi}$, where the dot represents derivation with respect to coordinate time. The perturbed field equations are as follows \cite{Gambino2024}: the (00) component,
\begin{align}
 &\,  6(1+\aB)H\dot{\Psi} + (6-\aK+12\aB)H^2\Phi - 2(1+\aH)\frac{(D^2+3{\mathcal{K}})}{a^2}\Psi + (\aK-6\aB)H^2\dot{\pi} \nonumber \\
 &\, +6\left[(1+\aB)\dot{H}+\frac{\rho_{\rm m} + p_{\rm m}}{2M^2}+(1-\aB)\frac{\mathcal{K}}{a^2} -\frac{1}{3}(\aH-\aB)\frac{(D^2+3{{ \mathcal{K}}})}{a^2}\right]H\pi = 
 -\frac{\delta\rho_{\rm m}}{M^2} \,;
\label{eq:00Newt}
\end{align}
the (0i) components,
\begin{equation}
   2\dot{\Psi} + 2(1+\aB)H\Phi - 2H\aB\dot{\pi} + \left(2\dot{H}+\frac{\rho_{\rm m}+p_{\rm m}}{M^2}\right)\pi = -\frac{(\rho_{\rm m}+p_{\rm m})v_{\rm m}}{M^2} \,;
\label{eq:0iNewt}
\end{equation}
the traceless part of the (ij) components,
\begin{equation}
     (1+\aH)\Phi - (1+\aT)\Psi + (\aM - \aT)H\pi - \aH\dot{\pi} = -\frac{\sigma_{\rm m}}{M^2} \,;
\label{eq:tracelessNewt}
\end{equation}
and, finally, the trace of the (ij) components,
\begin{align}  
    &\, 2\ddot{\Psi} + 2(3+\aM)H\dot{\Psi} + 2(1+\aB)H\dot{\Phi} \nonumber \\
    &\, + 2\left[\dot{H}-\frac{\rho_{\rm m} + p_{\rm m}}{2M^2}+(\aB H)\dot{} +(3+\aM)(1+\aB)H^2\right]\Phi \nonumber \\
    &\, - 2H\aB\ddot{\pi} + 2\left[\dot{H}+\frac{\rho_{\rm m} + p_{\rm m}}{2M^2}-(\aB H)\dot{}-(3+\alpha_{\rm M})\aB H^2\right]\dot{\pi} \nonumber \\
    &\, +2\left[(3+\aM)H\dot{H}+\frac{\dot{p}_{\rm m}}{2M^2}+\ddot{H}\right]\pi \nonumber \\
   &\,  -6\frac{{{\mathcal{K}}}}{a^2}[\dot{\pi}+(1+\aM)H\pi] = \frac{1}{M^2}\left(\delta p_{\rm m} + \frac{2}{3}\frac{(D^2+3\mathcal{K})}{a^2}\sigma_{\rm m}\right)\,.
    \label{eq:PsiNewt}
\end{align}
The equation of motion for $\pi$ is instead:
\begin{align} 
 &\, H^2\aK\ddot{\pi}  + \left\{\left[H^2(3+\aM)+\dot{H}\right]\aK + (H\aK)\dot{} \ \right\}H\dot{\pi}\\
 &\, + 6\left\{\left(\dot{H}+\frac{\rho_{\rm m} + p_{\rm m}}{2M^2}\right)\dot{H} + \dot{H}\aB\left[H^2(3+\aM)+\dot{H}\right] + H(\dot{H}\aB)\dot{} \ \right\}\pi \nonumber \\
 &\, + 2\frac{(D^2+3{\mathcal{K}})}{a^2}\left\{\dot{H} + \frac{\rho_{\rm m} + p_{\rm m}}{2M^2} + H^2[1+\aB(1+\aM)+\aT-(1+\aH)(1+\aM)] + (H(\aB-\aH))\dot{} \ \right\}\pi \nonumber \\ 
 &\, + 6H\aB\ddot{\Psi} + H^2(6\aB-\aK)\dot{\Phi} + 6\left[\dot{H}+\frac{\rho_{\rm m} + p_{\rm m}}{2M^2}+ H^2\aB(3+\aM) + (\aB H)\dot{}\right]\dot{\Psi} \nonumber \\
 &\, + \left[6\left(\dot{H}+\frac{\rho_{\rm m} + p_{\rm m}}{2M^2}\right)+H^2(6\aB-\aK)(3+\aM) +2(9\aB-\aK)\dot{H} +H(6\dot{\aB}-\dot{\aK})\right] H\Phi\nonumber \\
 &\, -\frac{6{\mathcal K}}{a^2}\left[H^2\aB(1-\aM)+(H\aB)\dot{}\ \right]\pi\nonumber \\
 &\, - 2\frac{(D^2+3{{ \mathcal{K}}})}{a^2}\left[\aH\dot{\Psi}+ [H(\alpha_{\rm M} + \aH(1+\alpha_{\rm M})-\aT) +\dot{\alpha}_{\rm H}]\Psi + H(\aH-\aB)\Phi\right]=0 \,.
  \label{eq:piNewt}
\end{align}
In all the above equations, the $\alpha$ functions are defined as in Eqs. (B.1)-(B.5) of \cite{Pace2019}. 
At the background level, the two Friedmann equations are as follows
\begin{equation}
    H^2 + \frac{\mathcal{K}}{a^2} = \frac{1}{3M^2}(\rho_{\rm m} + \rho_{\rm de})\,, \quad 
    \dot{H} - \frac{\mathcal{K}}{a^2} = -\frac{1}{2M^2}\left(\rho_{\rm m} + \rho_{\rm de} + P_{\rm m} + P_{\rm de}\right)\,,
\end{equation}
and the continuity equation for the dark sector reads
\begin{equation}
    \dot{\rho}_{\rm de} + 3H\left(\rho_{\rm de} + P_{\rm de}\right) = 3HM^2\aM\left(H^2 + \frac{\mathcal{K}}{a^2}\right)\,,
\end{equation}
or, equivalently,
\begin{equation}
    \dot{\rho}_{\rm de} = 3H(\rho_{\rm m} + P_{\rm m}) + 6HM^2\left(\dot{H} - \frac{\mathcal{K}}{a^2}\right) + 3HM^2\aM\left(H^2 + \frac{\mathcal{K}}{a^2}\right)\,.
\end{equation}

One can also express $\dot{P}_{\rm de}$ in terms of matter and geometry
\begin{equation}
    \dot{P}_{\rm de} = -\dot{P}_{\rm m} - M^2\left[2\ddot{H} + 2H\dot{H}(3+\aM) + 3H^3\aM - H(2-\aM)\frac{\mathcal{K}}{a^2}\right]\,.
\end{equation}
However, we usually want the dark sector to obey a standard continuity equation. We, therefore, define an effective fluid such that
\begin{equation}
    \rho_{\rm de} = \bar{M}^2\rho_{\rm de}^{\rm eff} + \left(\bar{M}^2-1\right)\rho_{\rm m}\,, \quad 
    P_{\rm de} = \bar{M}^2P_{\rm de}^{\rm eff} + \left(\bar{M}^2-1\right)P_{\rm m}\,,
\end{equation}
with $\bar{M}^2 = M^2/M_{\rm pl}^2$, where the Planck mass is $M_{\rm pl} = (8\pi G)^{-1/2}$.
In this case,
\begin{equation}
    H^2 + \frac{\mathcal{K}}{a^2} = \frac{1}{3M_{\rm pl}^2}(\rho_{\rm m} + \rho_{\rm de}^{\rm eff})\,, \quad 
    \dot{H} - \frac{\mathcal{K}}{a^2} = -\frac{1}{2M_{\rm pl}^2}\left(\rho_{\rm m} + \rho_{\rm de}^{\rm eff} + P_{\rm m} + P_{\rm de}^{\rm eff}\right)\,,
\end{equation}
and
\begin{equation}
    \dot{\rho}_{\rm de}^{\rm eff} + 3H\left(\rho_{\rm de}^{\rm eff} + P_{\rm de}^{\rm eff}\right) = 0\,.
\end{equation}

We can now write the effective dark energy density and pressure in terms of the functions $G_i$. To find the expressions for $\rho_{\rm de}^{\rm eff}$ and $P_{\rm de}^{\rm eff}$, we require that $H^2$ and $\dot{H}$ are the observable quantities that do not change whether we use one or the other definition for density and pressure. By doing so, we see that
\begin{equation}
    \rho_{\rm de} = \rho_{\rm de}^{\rm eff} + 3\left(M^2-M_{\rm pl}^2\right)\left(H^2+\frac{k}{a^2}\right)\,, \quad 
    P_{\rm de} = P_{\rm de}^{\rm eff} - \left(M^2 - M_{\rm pl}^2\right)\left(3H^2+2\dot{H}+\frac{k}{a^2}\right)\,,
\end{equation}
with
\begin{align}
    \rho_{\rm de}^{\rm eff} = &\, 2XG_{2,X} - G_2 - XG_{3,\phi} - 6H\dot{\phi}XG_{3,X} - 
                     6H^2(G_4 -4XG_{4,X}-4X^2G_{4,XX}) - \nonumber\\
                 &\, 6H\dot{\phi}(G_{4,\phi}+2XG_{4,\phi X}) + 
                     2H^3\dot{\phi}X(5G_{5,X}+2XG_{5,XX}) + 
                     3H^2X(3G_{5,\phi}+2XG_{5,\phi X}) + \nonumber\\
                 &\, 3H^2M_{\rm pl}^2\,,\\
    P_{\rm de}^{\rm eff} = &\, G_2 + X\left(2\ddot{\phi}G_{3,X}-G_{3,\phi}\right) + 2\left(3H^2+2\dot{H}\right)G_4 - 
                  4H\left(\dot{X}+3HX\right)G_{4,X} - \nonumber\\
              &\, 8X\left(\dot{H}G_{4,X}+H\dot{X}G_{4,XX}\right) + 
                  2\left(\ddot{\phi}+2H\dot{\phi}\right)G_{4,\phi} - 2XG_{4,\phi\phi} + 
                  4\left(\ddot{\phi}-2H\dot{\phi}\right)XG_{4,\phi X} - \nonumber\\
              &\, 2X\left(2H^3\dot{\phi}+2H\dot{H}\dot{\phi}+3H^2\ddot{\phi}\right)G_{5,X} - 
                  4H^2\ddot{\phi}X^2G_{5,XX} - 2HX\left(\dot{X}-HX\right)G_{5,\phi X} - \nonumber\\
              &\, \left[2\left(HX\right)^{\hbox{$\cdot$}}+3H^2X\right]G_{5,\phi} - 2H\dot{\phi}XG_{5,\phi\phi} -
                  \left(3H^2+2\dot{H}\right)M_{\rm pl}^2\,.
\end{align}

To apply the QSA and use the previous equations, we need to leave the configuration space. We cannot perform a simple Fourier transform as $D^2$ does not become $k^2$, but a differential operator in momentum space. We expand our functions in the harmonics $\Psi(\vec{x}) = \sum \Psi_q Q_{qlm}$ such that $D^2 Q_{qlm} = -(q^2-3\mathcal{K})Q_{qlm}$. Thus, terms like $D^2+3\mathcal{K}$ simply become $-q^2$. Hence, going to momentum space, the field equations (\ref{eq:00Newt})-(\ref{eq:PsiNewt}) and the equation of motion for $\pi$, Eq. (\ref{eq:piNewt}), become the following:
 \begin{align}
 &\,  6(1+\aB)H\dot{\Psi} + (6-\aK+12\aB)H^2\Phi + 2(1+\aH)\frac{q^2}{a^2}\Psi + (\aK-6\aB)H^2\dot{\pi} \nonumber \\
 &\, +6\left[(1+\aB)\dot{H}+\frac{\rho_{\rm m} + p_{\rm m}}{2M^2}+(1-\aB)\frac{\mathcal{K}}{a^2} + \frac{1}{3}(\aH-\aB)\frac{q^2}{a^2}\right]H\pi = 
 -\frac{\delta\rho_{\rm m}}{M^2} \,;
\label{eq:00Newt1}
\end{align}
\begin{equation}
   2\dot{\Psi} + 2(1+\aB)H\Phi - 2H\aB\dot{\pi} + \left(2\dot{H}+\frac{\rho_{\rm m}+p_{\rm m}}{M^2}\right)\pi = -\frac{(\rho_{\rm m}+p_{\rm m})v_{\rm m}}{M^2} \,;
\label{eq:0iNewt1}
\end{equation}
\begin{equation}
     (1+\aH)\Phi - (1+\aT)\Psi + (\aM - \aT)H\pi - \aH\dot{\pi} = -\frac{\sigma_{\rm m}}{M^2} \,;
\label{eq:tracelessNewt1}
\end{equation}
\begin{align}  
    &\, 2\ddot{\Psi} + 2(3+\aM)H\dot{\Psi} + 2(1+\aB)H\dot{\Phi} \nonumber \\
    &\, + 2\left[\dot{H}-\frac{\rho_{\rm m} + p_{\rm m}}{2M^2}+(\aB H)\dot{} +(3+\aM)(1+\aB)H^2\right]\Phi \nonumber \\
    &\, - 2H\aB\ddot{\pi} + 2\left[\dot{H}+\frac{\rho_{\rm m} + p_{\rm m}}{2M^2}-(\aB H)\dot{}-(3+\alpha_{\rm M})\aB H^2\right]\dot{\pi} \nonumber \\
    &\, +2\left[(3+\aM)H\dot{H}+\frac{\dot{p}_{\rm m}}{2M^2}+\ddot{H}\right]\pi \nonumber \\
   &\,  -6\frac{{{\mathcal{K}}}}{a^2}[\dot{\pi}+(1+\aM)H\pi] = \frac{1}{M^2}\left(\delta p_{\rm m} - \frac{2}{3}\frac{q^2}{a^2}\sigma_{\rm m}\right)\,;
    \label{eq:PsiNewt1}
\end{align}
\begin{align}
 &\, H^2\aK\ddot{\pi}  + \left\{\left[H^2(3+\aM)+\dot{H}\right]\aK + (H\aK)\dot{} \ \right\}H\dot{\pi}\\
 &\, + 6\left\{\left(\dot{H}+\frac{\rho_{\rm m} + p_{\rm m}}{2M^2}\right)\dot{H} + \dot{H}\aB\left[H^2(3+\aM)+\dot{H}\right] + H(\dot{H}\aB)\dot{} \ \right\}\pi \nonumber \\
 &\, - 2\frac{q^2}{a^2}\left\{\dot{H} + \frac{\rho_{\rm m} + p_{\rm m}}{2M^2} + H^2[1+\aB(1+\aM)+\aT-(1+\aH)(1+\aM)] + (H(\aB-\aH))\dot{} \ \right\}\pi \nonumber \\ 
 &\, + 6H\aB\ddot{\Psi} + H^2(6\aB-\aK)\dot{\Phi} + 6\left[\dot{H}+\frac{\rho_{\rm m} + p_{\rm m}}{2M^2}+ H^2\aB(3+\aM) + (\aB H)\dot{}\right]\dot{\Psi} \nonumber \\
 &\, + \left[6\left(\dot{H}+\frac{\rho_{\rm m} + p_{\rm m}}{2M^2}\right)+H^2(6\aB-\aK)(3+\aM) +2(9\aB-\aK)\dot{H} +H(6\dot{\aB}-\dot{\aK})\right] H\Phi\nonumber \\
 &\, -\frac{6{\mathcal K}}{a^2}\left[H^2\aB(1-\aM)+(H\aB)\dot{}\ \right]\pi\nonumber \\
 &\, + 2\frac{q^2}{a^2}\left[\aH\dot{\Psi}+ [H(\alpha_{\rm M} + \aH(1+\alpha_{\rm M})-\aT) +\dot{\alpha}_{\rm H}]\Psi + H(\aH-\aB)\Phi\right]=0 \,.
  \label{eq:piNewt1}
\end{align}
In these equations, $v_m$ represents the velocity potential of the matter content and $\sigma_m$ its anisotropic stress component, as they appear inside the $0i$ and $ij$ components of the standard perfect fluid stress-energy tensor.
To make these expressions more compact, we write them as in \cite{Pace2021}, and we denote $(\rho_{\rm m}+p_{\rm m})v_{\rm m} = q_{\rm m}$ and set $\aH = 0$,
\begin{align}
 &\,  C_{\dot{\Psi}}^{00}\dot{\Psi} + C_{\Phi}^{00}\Phi + C_{\Psi}^{00}\mathrm{Q}^2\Psi + C_{\dot{\pi}}^{00}\dot{\pi} + \left(C_{\pi}^{00} + C_{\pi2}^{00}\mathrm{Q}^2\right)\pi = 
 -\frac{\delta\rho_{\rm m}}{M^2} \,,\\
 &\, C_{\dot{\Psi}}^{0i}\dot{\Psi} + C_{\Phi}^{0i}\Phi + C_{\dot{\pi}}^{0i}\dot{\pi} + C_{\pi}^{0i}\pi = -\frac{q_{\rm m}}{M^2}\,, \\
 &\, C_{\Phi}^{ij,{\rm trl}}\Phi + C_{\Psi}^{ij,{\rm trl}}\Psi + C_{\pi}^{ij,{\rm trl}}\pi = -\frac{\sigma_{\rm m}}{M^2}\,, \\
 &\, C_{\ddot{\Psi}}^{ij,{\rm tr}}\ddot{\Psi} + C_{\dot{\Psi}}^{ij,{\rm tr}}\dot{\Psi} + C_{\dot{\Phi}}^{ij,{\rm tr}}\dot{\Phi} + C_{\Phi}^{ij,{\rm tr}}\Phi + 
     C_{\ddot{\pi}}^{ij,{\rm tr}}\ddot{\pi} + C_{\dot{\pi}}^{ij,{\rm tr}}\dot{\pi} + C_{\pi}^{ij,{\rm tr}}\pi - 6\frac{{{\mathcal{K}}}}{a^2}[\dot{\pi}+(1+\aM)H\pi] = \nonumber \\
 &\, \qquad\qquad    \frac{1}{M^2}\left(\delta p_{\rm m} - \frac{2}{3}H^2\mathrm{Q}^2\sigma_{\rm m}\right)\,, \\
 &\, C_{\ddot{\pi}}^{\pi}\ddot{\pi} + C_{\dot{\pi}}^{\pi}\dot{\pi} + \left(C_{\pi}^{\pi} + C_{\pi 2}^{\pi}\mathrm{Q}^2\right)\pi + C_{\ddot{\Psi}}^{\pi}\ddot{\Psi} + C_{\dot{\Phi}}^{\pi}\dot{\Phi} + C_{\dot{\Psi}}^{\pi}\dot{\Psi}  + \left(C_{\Phi}^{\pi} + C_{\Phi2}^{\pi}\mathrm{Q}^2\right)\Phi + C_{\Psi 2}^{\pi}\mathrm{Q}^2\Psi\nonumber \\
 &\, -\frac{6{\mathcal K}}{a^2}\left[H^2\aB(1-\aM)+(H\aB)\dot{}\ \right]\pi = 0 \,.
\end{align}
Here we have defined a set of coefficients $C^a_X$, where $a$ represents the equation in which a given coefficient appears (respectively $00$, $0i$, traceless $ij$ and trace $ij$ components of the field equations, and $\pi$ for the KG equation), and $X$ represents the term multiplying the coefficient (either $\Phi$, $\Psi$, $\pi$ or their first and second order derivatives).
The scale is given by $\mathrm{Q}^2 = \left(\frac{q}{aH}\right)^2$. The application of QSA implies that $\mathrm{Q} \gg 1$ and that we neglect the time derivatives of the potential and of the scalar field. We therefore only keep terms with $\mathrm{Q}^2$ and further neglect terms where the spatial curvature $\mathcal{K}$ appears. This quantity represents the background curvature and from the first Friedmann equation one can see that it is of the order of $H^2$, exactly as the terms we are neglecting. Combining the $00$-equation with the $0i$ equations in a gauge-invariant way and solving for $\sigma_{\rm m}$ from the traceless equation and placing it in the trace equation, we arrive at the following algebraic system
\begin{align}
 C_{\Psi}^{00}\mathrm{Q}^2\Psi + C_{\pi2}^{00}\mathrm{Q}^2\pi = &\, -\frac{\rho_{\rm m}\Delta_{\rm m}}{M^2} \,,\\
 C_{\Phi}^{ij,{\rm trl}}\Phi + C_{\Psi}^{ij,{\rm trl}}\Psi + C_{\pi}^{ij,{\rm trl}}\pi = &\, 0\,, \\
 C_{\Phi2}^{\pi}\mathrm{Q}^2\Phi + C_{\Psi 2}^{\pi}\mathrm{Q}^2\Psi + \left(C_{\pi}^{\pi} + C_{\pi 2}^{\pi}\mathrm{Q}^2\right)\pi = &\, 0\,.
\end{align}

The solution of this system is given in Eqs. (A.6) -- (A.9) in \cite{Pace2021}, once we replace $\mathrm{K} \to \mathrm{Q}$. We consider only the limit for $\mathrm{Q} \to \infty$ (see discussion there), hence we have the standard results without curvature. 

\bibliography{biblio}

@article{Matsakos,
    author = "Matsakos, Titos and Diaferio, Antonaldo",
    title = "{Dynamics of galaxies and clusters in refracted gravity}",
    eprint = "1603.04943",
    archivePrefix = "arXiv",
    primaryClass = "astro-ph.GA",
    month = "3",
    year = "2016"
}

@article{Sanna,
    author = "Sanna, Andrea Pierfrancesco and Matsakos, Titos and Diaferio, Antonaldo",
    title = "{Covariant formulation of refracted gravity}",
    eprint = "2109.11217",
    archivePrefix = "arXiv",
    primaryClass = "astro-ph.CO",
    doi = "10.1051/0004-6361/202243553",
    journal = "Astron. Astrophys.",
    volume = "674",
    pages = "A209",
    year = "2023"
}

@ARTICLE{Pizzuti,
       author = {{Pizzuti}, L. and {Fantoccoli}, F. and {Broccolato}, V. and {Biviano}, A. and {Diaferio}, A.},
        title = "{Testing refracted gravity with the kinematics of galaxy clusters}",
      journal = {\aap},
         year = 2025,
        month = jun,
       volume = {698},
          eid = {A83},
        pages = {A83},
          doi = {10.1051/0004-6361/202452739},
archivePrefix = {arXiv},
       eprint = {2410.19698},
 primaryClass = {astro-ph.CO},
       adsurl = {https://ui.adsabs.harvard.edu/abs/2025A&A...698A..83P}
}

@article{Quiros,
    author = "Quiros, Israel and Garc\'\i{}a-Salcedo, Ricardo and Gonzalez, Tame and Horta-Rangel, F. Antonio and Saavedra, Joel",
    title = "{Brans\textendash{}Dicke Galileon and the variational principle}",
    eprint = "1605.00326",
    archivePrefix = "arXiv",
    primaryClass = "gr-qc",
    doi = "10.1088/0143-0807/37/5/055605",
    journal = "Eur. J. Phys.",
    volume = "37",
    number = "5",
    pages = "055605",
    year = "2016"
}

@article{Kobayashi:2019,
    author = "Kobayashi, Tsutomu",
    title = "{Horndeski theory and beyond: a review}",
    eprint = "1901.07183",
    archivePrefix = "arXiv",
    primaryClass = "gr-qc",
    reportNumber = "RUP-19-3",
    doi = "10.1088/1361-6633/ab2429",
    journal = "Rept. Prog. Phys.",
    volume = "82",
    number = "8",
    pages = "086901",
    year = "2019"
}

@article{Horndeski:1974wa,
    author = "Horndeski, Gregory Walter",
    title = "{Second-order scalar-tensor field equations in a four-dimensional space}",
    doi = "10.1007/BF01807638",
    journal = "Int. J. Theor. Phys.",
    volume = "10",
    pages = "363--384",
    year = "1974"
}

@article{Kobayashi:2011nu,
    author = "Kobayashi, Tsutomu and Yamaguchi, Masahide and Yokoyama, Jun'ichi",
    title = "{Generalized G-inflation: Inflation with the most general second-order field equations}",
    eprint = "1105.5723",
    archivePrefix = "arXiv",
    primaryClass = "hep-th",
    reportNumber = "KUNS-2339, RESCEU-9-11",
    doi = "10.1143/PTP.126.511",
    journal = "Prog. Theor. Phys.",
    volume = "126",
    pages = "511--529",
    year = "2011"
}

@article{Brans:1961sx,
    author = "Brans, C. and Dicke, R. H.",
    editor = "Hsu, Jong-Ping and Fine, D.",
    title = "{Mach's principle and a relativistic theory of gravitation}",
    doi = "10.1103/PhysRev.124.925",
    journal = "Phys. Rev.",
    volume = "124",
    pages = "925--935",
    year = "1961"
}

@article{Yu:2017iju,
    author = "Yu, Hai and Ratra, Bharat and Wang, Fa-Yin",
    title = "{Hubble Parameter and Baryon Acoustic Oscillation Measurement Constraints on the Hubble Constant, the Deviation from the Spatially Flat \ensuremath{\Lambda}CDM Model, the Deceleration\textendash{}Acceleration Transition Redshift, and Spatial Curvature}",
    eprint = "1711.03437",
    archivePrefix = "arXiv",
    primaryClass = "astro-ph.CO",
    doi = "10.3847/1538-4357/aab0a2",
    journal = "Astrophys. J.",
    volume = "856",
    number = "1",
    pages = "3",
    year = "2018"
}

@article{Bertacca:2008uf,
    author = "Bertacca, Daniele and Bartolo, Nicola and Diaferio, Antonaldo and Matarrese, Sabino",
    title = "{How the Scalar Field of Unified Dark Matter Models Can Cluster}",
    eprint = "0807.1020",
    archivePrefix = "arXiv",
    primaryClass = "astro-ph",
    doi = "10.1088/1475-7516/2008/10/023",
    journal = "JCAP",
    volume = "10",
    pages = "023",
    year = "2008"
}

@article{Cesare:2020pyz,
    author = "Cesare, Valentina and Diaferio, Antonaldo and Matsakos, Titos and Angus, Garry",
    title = "{Dynamics of DiskMass Survey galaxies in refracted gravity}",
    eprint = "2003.07377",
    archivePrefix = "arXiv",
    primaryClass = "astro-ph.GA",
    doi = "10.1051/0004-6361/201935950",
    journal = "Astron. Astrophys.",
    volume = "637",
    pages = "A70",
    year = "2020"
}

@article{McGaugh:2016leg,
    author = "McGaugh, Stacy and Lelli, Federico and Schombert, Jim",
    title = "{Radial Acceleration Relation in Rotationally Supported Galaxies}",
    eprint = "1609.05917",
    archivePrefix = "arXiv",
    primaryClass = "astro-ph.GA",
    doi = "10.1103/PhysRevLett.117.201101",
    journal = "Phys. Rev. Lett.",
    volume = "117",
    number = "20",
    pages = "201101",
    year = "2016"
}

@article{Lelli:2016cui,
    author = "Lelli, Federico and McGaugh, Stacy S. and Schombert, James M. and Pawlowski, Marcel S.",
    title = "{One Law to Rule Them All: The Radial Acceleration Relation of Galaxies}",
    eprint = "1610.08981",
    archivePrefix = "arXiv",
    primaryClass = "astro-ph.GA",
    doi = "10.3847/1538-4357/836/2/152",
    journal = "Astrophys. J.",
    volume = "836",
    number = "2",
    pages = "152",
    year = "2017"
}

@article{Cesare:2022khd,
    author = "Cesare, V. and Diaferio, A. and Matsakos, T.",
    title = "{The dynamics of three nearby E0 galaxies in refracted gravity}",
    eprint = "2102.12499",
    archivePrefix = "arXiv",
    doi = "10.1051/0004-6361/202140651",
    journal = "Astron. Astrophys.",
    volume = "657",
    pages = "A133",
    year = "2022"
}

@article{Mathews:2019hbi,
    author = "Mathews, G. J. and Kedia, A. and Sasankan, N. and Kusakabe, M. and Luo, Y. and Kajino, T. and Yamazaki, D. and Makki, T. and Eid, M. El",
    editor = "Kawabata, T. and others",
    title = "{Cosmological Solutions to the Lithium Problem}",
    eprint = "1909.01245",
    archivePrefix = "arXiv",
    primaryClass = "astro-ph.CO",
    doi = "10.7566/JPSCP.31.011033",
    journal = "JPS Conf. Proc.",
    volume = "31",
    pages = "011033",
    year = "2020"
}

@article{Weinberg:1988cp,
    author = "Weinberg, Steven",
    editor = "Hsu, Jong-Ping and Fine, D.",
    title = "{The Cosmological Constant Problem}",
    reportNumber = "UTTG-12-88",
    doi = "10.1103/RevModPhys.61.1",
    journal = "Rev. Mod. Phys.",
    volume = "61",
    pages = "1--23",
    year = "1989"
}

@article{Padilla:2015aaa,
    author = "Padilla, Antonio",
    title = "{Lectures on the Cosmological Constant Problem}",
    eprint = "1502.05296",
    archivePrefix = "arXiv",
    primaryClass = "hep-th",
    month = "2",
    year = "2015"
}

@article{Moore:1999nt,
    author = "Moore, B. and Ghigna, S. and Governato, F. and Lake, G. and Quinn, Thomas R. and Stadel, J. and Tozzi, P.",
    title = "{Dark matter substructure within galactic halos}",
    eprint = "astro-ph/9907411",
    archivePrefix = "arXiv",
    doi = "10.1086/312287",
    journal = "Astrophys. J. Lett.",
    volume = "524",
    pages = "L19--L22",
    year = "1999"
}

@article{Famaey:2011kh,
    author = "Famaey, Benoit and McGaugh, Stacy",
    title = "{Modified Newtonian Dynamics (MOND): Observational Phenomenology and Relativistic Extensions}",
    eprint = "1112.3960",
    archivePrefix = "arXiv",
    primaryClass = "astro-ph.CO",
    doi = "10.12942/lrr-2012-10",
    journal = "Living Rev. Rel.",
    volume = "15",
    pages = "10",
    year = "2012"
}

@article{deMartino:2020gfi,
    author = "de Martino, Ivan and Chakrabarty, Sankha S. and Cesare, Valentina and Gallo, Arianna and Ostorero, Luisa and Diaferio, Antonaldo",
    title = "{Dark matters on the scale of galaxies}",
    eprint = "2007.15539",
    archivePrefix = "arXiv",
    primaryClass = "astro-ph.CO",
    doi = "10.3390/universe6080107",
    journal = "Universe",
    volume = "6",
    number = "8",
    pages = "107",
    year = "2020"
}

@article{Klypin:1999uc,
    author = "Klypin, Anatoly A. and Kravtsov, Andrey V. and Valenzuela, Octavio and Prada, Francisco",
    title = "{Where are the missing Galactic satellites?}",
    eprint = "astro-ph/9901240",
    archivePrefix = "arXiv",
    doi = "10.1086/307643",
    journal = "Astrophys. J.",
    volume = "522",
    pages = "82--92",
    year = "1999"
}

@article{Bullock:2017xww,
    author = "Bullock, James S. and Boylan-Kolchin, Michael",
    title = "{Small-Scale Challenges to the $\Lambda$CDM Paradigm}",
    eprint = "1707.04256",
    archivePrefix = "arXiv",
    primaryClass = "astro-ph.CO",
    doi = "10.1146/annurev-astro-091916-055313",
    journal = "Ann. Rev. Astron. Astrophys.",
    volume = "55",
    pages = "343--387",
    year = "2017"
}

@article{McGaugh:1998tq,
    author = "McGaugh, Stacy S. and de Blok, W. J. G.",
    title = "{Testing the dark matter hypothesis with low surface brightness galaxies and other evidence}",
    eprint = "astro-ph/9801123",
    archivePrefix = "arXiv",
    doi = "10.1086/305612",
    journal = "Astrophys. J.",
    volume = "499",
    pages = "41",
    year = "1998"
}

@article{Boylan-Kolchin:2011qkt,
    author = "Boylan-Kolchin, Michael and Bullock, James S. and Kaplinghat, Manoj",
    title = "{Too big to fail? The puzzling darkness of massive Milky Way subhaloes}",
    eprint = "1103.0007",
    archivePrefix = "arXiv",
    primaryClass = "astro-ph.CO",
    doi = "10.1111/j.1745-3933.2011.01074.x",
    journal = "Mon. Not. Roy. Astron. Soc.",
    volume = "415",
    pages = "L40",
    year = "2011"
}

@article{McGaugh:2000sr,
    author = "McGaugh, Stacy S. and Schombert, Jim M. and Bothun, Greg D. and de Blok, W. J. G.",
    title = "{The Baryonic Tully-Fisher relation}",
    eprint = "astro-ph/0003001",
    archivePrefix = "arXiv",
    doi = "10.1086/312628",
    journal = "Astrophys. J. Lett.",
    volume = "533",
    pages = "L99--L102",
    year = "2000"
}

@article{Pawlowski:2014una,
    author = "Pawlowski, Marcel S. and others",
    title = "{Co-orbiting satellite galaxy structures are still in conflict with the distribution of primordial dwarf galaxies}",
    eprint = "1406.1799",
    archivePrefix = "arXiv",
    primaryClass = "astro-ph.GA",
    doi = "10.1093/mnras/stu1005",
    journal = "Mon. Not. Roy. Astron. Soc.",
    volume = "442",
    number = "3",
    pages = "2362--2380",
    year = "2014"
}

@article{Gentile:2004tb,
    author = "Gentile, Gianfranco and Salucci, P. and Klein, U. and Vergani, D. and Kalberla, P.",
    title = "{The Cored distribution of dark matter in spiral galaxies}",
    eprint = "astro-ph/0403154",
    archivePrefix = "arXiv",
    doi = "10.1111/j.1365-2966.2004.07836.x",
    journal = "Mon. Not. Roy. Astron. Soc.",
    volume = "351",
    pages = "903",
    year = "2004"
}

@article{Milgrom:1983ca,
    author = "Milgrom, M.",
    title = "{A Modification of the Newtonian dynamics as a possible alternative to the hidden mass hypothesis}",
    doi = "10.1086/161130",
    journal = "Astrophys. J.",
    volume = "270",
    pages = "365--370",
    year = "1983"
}

@article{Milgrom:1983pn,
    author = "Milgrom, M.",
    title = "{A Modification of the Newtonian dynamics: Implications for galaxies}",
    doi = "10.1086/161131",
    journal = "Astrophys. J.",
    volume = "270",
    pages = "371--383",
    year = "1983"
}

@article{Milgrom:1983zz,
    author = "Milgrom, M.",
    title = "{A modification of the Newtonian dynamics: implications for galaxy systems}",
    doi = "10.1086/161132",
    journal = "Astrophys. J.",
    volume = "270",
    pages = "384--389",
    year = "1983"
}

@article{Skordis:2020eui,
    author = "Skordis, Constantinos and Zlosnik, Tom",
    title = "{New Relativistic Theory for Modified Newtonian Dynamics}",
    eprint = "2007.00082",
    archivePrefix = "arXiv",
    primaryClass = "astro-ph.CO",
    doi = "10.1103/PhysRevLett.127.161302",
    journal = "Phys. Rev. Lett.",
    volume = "127",
    number = "16",
    pages = "161302",
    year = "2021"
}

@article{Bertacca:2010ct,
    author = "Bertacca, Daniele and Bartolo, Nicola and Matarrese, Sabino",
    title = "{Unified Dark Matter Scalar Field Models}",
    eprint = "1008.0614",
    archivePrefix = "arXiv",
    primaryClass = "astro-ph.CO",
    doi = "10.1155/2010/904379",
    journal = "Adv. Astron.",
    volume = "2010",
    pages = "904379",
    year = "2010"
}

@article{DESI:2025zgx,
    author = "Abdul Karim, M. and others",
    collaboration = "DESI",
    title = "{DESI DR2 Results II: Measurements of Baryon Acoustic Oscillations and Cosmological Constraints}",
    eprint = "2503.14738",
    archivePrefix = "arXiv",
    primaryClass = "astro-ph.CO",
    reportNumber = "FERMILAB-PUB-25-0169-PPD",
    month = "3",
    year = "2025"
}

@article{CosmoVerse:2025txj,
    author = "Di Valentino, Eleonora and others",
    collaboration = "CosmoVerse",
    title = "{The CosmoVerse White Paper: Addressing observational tensions in cosmology with systematics and fundamental physics}",
    eprint = "2504.01669",
    archivePrefix = "arXiv",
    primaryClass = "astro-ph.CO",
    month = "4",
    year = "2025"
}

@article{Planck:2018vyg,
    author = "Aghanim, N. and others",
    collaboration = "Planck",
    title = "{Planck 2018 results. VI. Cosmological parameters}",
    eprint = "1807.06209",
    archivePrefix = "arXiv",
    primaryClass = "astro-ph.CO",
    doi = "10.1051/0004-6361/201833910",
    journal = "Astron. Astrophys.",
    volume = "641",
    pages = "A6",
    year = "2020",
    note = "[Erratum: Astron.Astrophys. 652, C4 (2021)]"
}

@ARTICLE{Silvestri2013,
       author = {{Silvestri}, Alessandra and {Pogosian}, Levon and {Buniy}, Roman V.},
        title = "{Practical approach to cosmological perturbations in modified gravity}",
      journal = {\prd},
         year = 2013,
        month = may,
       volume = {87},
       number = {10},
          eid = {104015},
        pages = {104015},
          doi = {10.1103/PhysRevD.87.104015},
archivePrefix = {arXiv},
       eprint = {1302.1193},
 primaryClass = {astro-ph.CO},
       adsurl = {https://ui.adsabs.harvard.edu/abs/2013PhRvD..87j4015S}
}

@ARTICLE{DeFelice2012,
       author = {{De Felice}, Antonio and {Tsujikawa}, Shinji},
        title = "{Conditions for the cosmological viability of the most general scalar-tensor theories and their applications to extended Galileon dark energy models}",
      journal = {\jcap},
         year = 2012,
        month = feb,
       volume = {2012},
       number = {2},
          eid = {007},
        pages = {007},
          doi = {10.1088/1475-7516/2012/02/007},
archivePrefix = {arXiv},
       eprint = {1110.3878},
 primaryClass = {gr-qc},
       adsurl = {https://ui.adsabs.harvard.edu/abs/2012JCAP...02..007D}
}

@ARTICLE{Pace2019,
       author = {{Pace}, Francesco and {Battye}, Richard A. and {Bolliet}, Boris and {Trinh}, Damien},
        title = "{Dark sector evolution in Horndeski models}",
      journal = {\jcap},
         year = 2019,
        month = sep,
       volume = {2019},
       number = {9},
          eid = {018},
        pages = {018},
          doi = {10.1088/1475-7516/2019/09/018},
archivePrefix = {arXiv},
       eprint = {1905.06795},
 primaryClass = {astro-ph.CO},
       adsurl = {https://ui.adsabs.harvard.edu/abs/2019JCAP...09..018P}
}

@ARTICLE{Pace2021,
       author = {{Pace}, Francesco and {Battye}, Richard A. and {Bellini}, Emilio and {Lombriser}, Lucas and {Vernizzi}, Filippo and {Bolliet}, Boris},
        title = "{Comparison of different approaches to the quasi-static approximation in Horndeski models}",
      journal = {\jcap},
         year = 2021,
        month = jun,
       volume = {2021},
       number = {6},
          eid = {017},
        pages = {017},
          doi = {10.1088/1475-7516/2021/06/017},
archivePrefix = {arXiv},
       eprint = {2011.05713},
 primaryClass = {astro-ph.CO},
       adsurl = {https://ui.adsabs.harvard.edu/abs/2021JCAP...06..017P}
}

@article{Cyburt:2015mya,
    author = "Cyburt, Richard H. and Fields, Brian D. and Olive, Keith A. and Yeh, Tsung-Han",
    title = "{Big Bang Nucleosynthesis: 2015}",
    eprint = "1505.01076",
    archivePrefix = "arXiv",
    primaryClass = "astro-ph.CO",
    reportNumber = "UMN-TH-3432-15, FTPI-MINN-15-19",
    doi = "10.1103/RevModPhys.88.015004",
    journal = "Rev. Mod. Phys.",
    volume = "88",
    pages = "015004",
    year = "2016"
}

@article{Fields:2019pfx,
    author = "Fields, Brian D. and Olive, Keith A. and Yeh, Tsung-Han and Young, Charles",
    title = "{Big-Bang Nucleosynthesis after Planck}",
    eprint = "1912.01132",
    archivePrefix = "arXiv",
    primaryClass = "astro-ph.CO",
    reportNumber = "UMN--TH--3902/19, FTPI--MINN--19/25",
    doi = "10.1088/1475-7516/2020/03/010",
    journal = "JCAP",
    volume = "03",
    pages = "010",
    year = "2020",
    note = "[Erratum: JCAP 11, E02 (2020)]"
}

@article{Planck:2019nip,
    author = "Aghanim, N. and others",
    collaboration = "Planck",
    title = "{Planck 2018 results. V. CMB power spectra and likelihoods}",
    eprint = "1907.12875",
    archivePrefix = "arXiv",
    primaryClass = "astro-ph.CO",
    doi = "10.1051/0004-6361/201936386",
    journal = "Astron. Astrophys.",
    volume = "641",
    pages = "A5",
    year = "2020"
}

@article{DelPopolo:2013qba,
    author = "Del Popolo, A.",
    title = "{Nonbaryonic Dark Matter in Cosmology}",
    eprint = "1305.0456",
    archivePrefix = "arXiv",
    primaryClass = "astro-ph.CO",
    doi = "10.1142/S0218271814300055",
    journal = "Int. J. Mod. Phys. D",
    volume = "23",
    pages = "1430005",
    year = "2014"
}

@article{Clowe:2006eq,
    author = "Clowe, Douglas and Bradac, Marusa and Gonzalez, Anthony H. and Markevitch, Maxim and Randall, Scott W. and Jones, Christine and Zaritsky, Dennis",
    title = "{A direct empirical proof of the existence of dark matter}",
    eprint = "astro-ph/0608407",
    archivePrefix = "arXiv",
    reportNumber = "SLAC-PUB-12078",
    doi = "10.1086/508162",
    journal = "Astrophys. J. Lett.",
    volume = "648",
    pages = "L109--L113",
    year = "2006"
}

@article{Iocco:2015xga,
    author = "Iocco, Fabio and Pato, Miguel and Bertone, Gianfranco",
    title = "{Evidence for dark matter in the inner Milky Way}",
    eprint = "1502.03821",
    archivePrefix = "arXiv",
    primaryClass = "astro-ph.GA",
    doi = "10.1038/nphys3237",
    journal = "Nature Phys.",
    volume = "11",
    pages = "245--248",
    year = "2015"
}

@article{Hu:2025fsz,
    author = "Hu, Jianping and Jia, Xuandong and Gao, DaoHong and Gao, Jiaze and Gao, Baoquan and Wang, Fayin",
    title = "{Constraints on transition redshift utilizing the latest H(z) measurements and comments on the Hubble tension}",
    eprint = "2508.05389",
    archivePrefix = "arXiv",
    primaryClass = "astro-ph.CO",
    month = "8",
    year = "2025"
}

@article{Hu:2024big,
    author = "Hu, J. P. and Jia, X. D. and Hu, J. and Wang, F. Y.",
    title = "{Hints of New Physics for the Hubble Tension: Violation of Cosmological Principle}",
    eprint = "2410.06450",
    archivePrefix = "arXiv",
    primaryClass = "astro-ph.CO",
    doi = "10.3847/2041-8213/ad85cf",
    journal = "Astrophys. J. Lett.",
    volume = "975",
    number = "2",
    pages = "L36",
    year = "2024"
}

@article{Moresco:2022phi,
    author = "Moresco, Michele and others",
    title = "{Unveiling the Universe with emerging cosmological probes}",
    eprint = "2201.07241",
    archivePrefix = "arXiv",
    primaryClass = "astro-ph.CO",
    doi = "10.1007/s41114-022-00040-z",
    journal = "Living Rev. Rel.",
    volume = "25",
    number = "1",
    pages = "6",
    year = "2022"
}

@article{Moresco:2023zys,
    author = "Moresco, Michele",
    title = "{Addressing the Hubble tension with cosmic chronometers}",
    eprint = "2307.09501",
    archivePrefix = "arXiv",
    primaryClass = "astro-ph.CO",
    month = "7",
    year = "2023"
}

@article{Yang:2024kdo,
    author = "Yang, Yuhang and Ren, Xin and Wang, Qingqing and Lu, Zhiyu and Zhang, Dongdong and Cai, Yi-Fu and Saridakis, Emmanuel N.",
    title = "{Quintom cosmology and modified gravity after DESI 2024}",
    eprint = "2404.19437",
    archivePrefix = "arXiv",
    primaryClass = "astro-ph.CO",
    doi = "10.1016/j.scib.2024.07.029",
    journal = "Sci. Bull.",
    volume = "69",
    pages = "2698--2704",
    year = "2024"
}

@misc{naidu2025cosmicmiracleremarkablyluminous,
      title={A Cosmic Miracle: A Remarkably Luminous Galaxy at $z_{\rm{spec}}=14.44$ Confirmed with JWST}, 
      author={Rohan P. Naidu and Pascal A. Oesch and Gabriel Brammer and Andrea Weibel and Yijia Li and Jorryt Matthee and John Chisholm and Clara L. Pollock and Kasper E. Heintz and Benjamin D. Johnson and Xuejian Shen and Raphael E. Hviding and Joel Leja and Sandro Tacchella and Arpita Ganguly and Callum Witten and Hakim Atek and Sirio Belli and Sownak Bose and Rychard Bouwens and Pratika Dayal and Roberto Decarli and Anna de Graaff and Yoshinobu Fudamoto and Emma Giovinazzo and Jenny E. Greene and Garth Illingworth and Akio K. Inoue and Sarah G. Kane and Ivo Labbe and Ecaterina Leonova and Rui Marques-Chaves and Romain A. Meyer and Erica J. Nelson and Guido Roberts-Borsani and Daniel Schaerer and Robert A. Simcoe and Mauro Stefanon and Yuma Sugahara and Sune Toft and Arjen van der Wel and Pieter van Dokkum and Fabian Walter and Darach Watson and John R. Weaver and Katherine E. Whitaker},
      year={2025},
      eprint={2505.11263},
      archivePrefix={arXiv},
      primaryClass={astro-ph.GA},
      url={https://arxiv.org/abs/2505.11263}, 
}

@article{M_Moresco_2012,
	title={Improved constraints on the expansion rate of the Universe up to z = 1.1 from the spectroscopic evolution of cosmic chronometers},
	volume={2012},
	ISSN={1475-7516},
	url={http://dx.doi.org/10.1088/1475-7516/2012/08/006},
	DOI={10.1088/1475-7516/2012/08/006},
	number={08},
	journal={Journal of Cosmology and Astroparticle Physics},
	publisher={IOP Publishing},
	author={M. Moresco and A. Cimatti and R. Jimenez and L. Pozzetti and G. Zamorani and M. Bolzonella and J. Dunlop and F. Lamareille and M. Mignoli and H. Pearce and P. Rosati and D. Stern and L. Verde and E. Zucca and C.M. Carollo and T. Contini and J.-P. Kneib and O. Le Fèvre and S.J. Lilly and V. Mainieri and A. Renzini and M. Scodeggio and I. Balestra and R. Gobat and R. McLure and S. Bardelli and A. Bongiorno and K. Caputi and O. Cucciati and S. de la Torre and L. de Ravel and P. Franzetti and B. Garilli and A. Iovino and P. Kampczyk and C. Knobel and K. Kovač and J.-F. Le Borgne and V. Le Brun and C. Maier and R. Pelló and Y. Peng and E. Perez-Montero and V. Presotto and J.D. Silverman and M. Tanaka and L.A.M. Tasca and L. Tresse and D. Vergani and O. Almaini and L. Barnes and R. Bordoloi and E. Bradshaw and A. Cappi and R. Chuter and M. Cirasuolo and G. Coppa and C. Diener and S. Foucaud and W. Hartley and M. Kamionkowski and A.M. Koekemoer and C. López-Sanjuan and H.J. McCracken and P. Nair and P. Oesch and A. Stanford and N. Welikala},
	year={2012},
	month=aug, pages={006–006} 
	}

@article{Zhang_2014,
title={Four new observational H(z) data from luminous red galaxies in the Sloan Digital Sky Survey data release seven},
DOI={10.1088/1674-4527/14/10/002},
journal={Research in Astronomy and Astrophysics},
volume={14},
issue={10},
author={Zhang, Cong and Zhang, Han and Yuan, Shuo and Liu, Siqi and Zhang, Tong-Jie and Sun, Yan-Chun},
year={2014}}

@article{Boylan-Kolchin_2023,
	author = {Boylan-Kolchin, Michael},
	date = {2023/06/01},
	doi = {10.1038/s41550-023-01937-7},
	id = {Boylan-Kolchin2023},
	isbn = {2397-3366},
	journal = {Nature Astronomy},
	number = {6},
	pages = {731--735},
	title = {Stress testing $\Lambda$CDM with high-redshift galaxy candidates},
	url = {https://doi.org/10.1038/s41550-023-01937-7},
	volume = {7},
	year = {2023}}

@article{Labbe_2023,
	author = {Labb{\'e}, Ivo and van Dokkum, Pieter and Nelson, Erica and Bezanson, Rachel and Suess, Katherine A. and Leja, Joel and Brammer, Gabriel and Whitaker, Katherine and Mathews, Elijah and Stefanon, Mauro and Wang, Bingjie},
	date = {2023/04/01},
	doi = {10.1038/s41586-023-05786-2},
	id = {Labb{\'e}2023},
	isbn = {1476-4687},
	journal = {Nature},
	number = {7956},
	pages = {266--269},
	title = {A population of red candidate massive galaxies \~{}600 Myr after the Big Bang},
	url = {https://doi.org/10.1038/s41586-023-05786-2},
	volume = {616},
	year = {2023}}

@article{Noeller2019,
  title = {Cosmological parameter constraints for Horndeski scalar-tensor gravity},
  author = {Noller, Johannes and Nicola, Andrina},
  journal = {Phys. Rev. D},
  volume = {99},
  issue = {10},
  pages = {103502},
  numpages = {19},
  year = {2019},
  month = {May},
  publisher = {American Physical Society},
  doi = {10.1103/PhysRevD.99.103502},
  url = {https://link.aps.org/doi/10.1103/PhysRevD.99.103502}
}

@article{Creminelli:2017sry,
    author = "Creminelli, Paolo and Vernizzi, Filippo",
    title = "{Dark Energy after GW170817 and GRB170817A}",
    eprint = "1710.05877",
    archivePrefix = "arXiv",
    primaryClass = "astro-ph.CO",
    doi = "10.1103/PhysRevLett.119.251302",
    journal = "Phys. Rev. Lett.",
    volume = "119",
    number = "25",
    pages = "251302",
    year = "2017"
}

@article{Sakstein:2017xjx,
    author = "Sakstein, Jeremy and Jain, Bhuvnesh",
    title = "{Implications of the Neutron Star Merger GW170817 for Cosmological Scalar-Tensor Theories}",
    eprint = "1710.05893",
    archivePrefix = "arXiv",
    primaryClass = "astro-ph.CO",
    doi = "10.1103/PhysRevLett.119.251303",
    journal = "Phys. Rev. Lett.",
    volume = "119",
    number = "25",
    pages = "251303",
    year = "2017"
}

@ARTICLE{Gervani:2026zqm,
       author = {{Gervani}, Lorenzo and {Diaferio}, Antonaldo and {Pace}, Francesco and {Sanna}, Andrea P.},
        title = "{Linear perturbation theory and structure formation in a Brans-Dicke theory of gravity without dark matter}",
      journal = {\jcap},
         year = 2026,
        month = jun,
       volume = {2026},
       number = {6},
          eid = {038},
        pages = {038},
          doi = {10.1088/1475-7516/2026/06/038},
archivePrefix = {arXiv},
       eprint = {2601.22937},
 primaryClass = {astro-ph.CO},
       adsurl = {https://ui.adsabs.harvard.edu/abs/2026JCAP...06..038G}
}

@article{SDSS:2001afu,
    author = "Dodelson, Scott and others",
    collaboration = "SDSS",
    title = "{The Three-Dimensional Power Spectrum from Angular Clustering of Galaxies in early SDSS Data}",
    eprint = "astro-ph/0107421",
    archivePrefix = "arXiv",
    reportNumber = "FERMILAB-PUB-01-271-A",
    doi = "10.1086/340225",
    journal = "Astrophys. J.",
    volume = "572",
    pages = "140--156",
    year = "2001"
}

@article{Bartelmann:1999yn,
    author = "Bartelmann, Matthias and Schneider, Peter",
    title = "{Weak gravitational lensing}",
    eprint = "astro-ph/9912508",
    archivePrefix = "arXiv",
    doi = "10.1016/S0370-1573(00)00082-X",
    journal = "Phys. Rept.",
    volume = "340",
    pages = "291--472",
    year = "2001"
}

@article{Kuhlen:2012ft,
    author = "Kuhlen, Michael and Vogelsberger, Mark and Angulo, Raul",
    title = "{Numerical Simulations of the Dark Universe: State of the Art and the Next Decade}",
    eprint = "1209.5745",
    archivePrefix = "arXiv",
    primaryClass = "astro-ph.CO",
    doi = "10.1016/j.dark.2012.10.002",
    journal = "Phys. Dark Univ.",
    volume = "1",
    pages = "50--93",
    year = "2012"
}

@article{Navarro:1996gj,
    author = "Navarro, Julio F. and Frenk, Carlos S. and White, Simon D. M.",
    title = "{A Universal density profile from hierarchical clustering}",
    eprint = "astro-ph/9611107",
    archivePrefix = "arXiv",
    doi = "10.1086/304888",
    journal = "Astrophys. J.",
    volume = "490",
    pages = "493--508",
    year = "1997"
}

@article{Salucci:2018hqu,
    author = "Salucci, Paolo",
    title = "{The distribution of dark matter in galaxies}",
    eprint = "1811.08843",
    archivePrefix = "arXiv",
    primaryClass = "astro-ph.GA",
    doi = "10.1007/s00159-018-0113-1",
    journal = "Astron. Astrophys. Rev.",
    volume = "27",
    number = "1",
    pages = "2",
    year = "2019"
}

@article{Salucci:2018eie,
    author = "Salucci, Paolo",
    title = "{Dark Matter in Galaxies: evidences and challenges}",
    eprint = "1807.08541",
    archivePrefix = "arXiv",
    primaryClass = "astro-ph.GA",
    doi = "10.1007/s10701-018-0209-5",
    journal = "Found. Phys.",
    volume = "48",
    number = "10",
    pages = "1517--1537",
    year = "2018"
}

@phdthesis{Wallisch:2018rzj,
    author = "Wallisch, Benjamin",
    title = "{Cosmological Probes of Light Relics}",
    eprint = "1810.02800",
    archivePrefix = "arXiv",
    primaryClass = "astro-ph.CO",
    doi = "10.17863/CAM.30368",
    school = "Cambridge U.",
    year = "2018"
}

@article{Riess_2024,
doi = {10.3847/1538-4357/ad8c21},
url = {https://doi.org/10.3847/1538-4357/ad8c21},
year = {2024},
month = {dec},
publisher = {The American Astronomical Society},
volume = {977},
number = {1},
pages = {120},
author = {Riess, Adam G. and Scolnic, Dan and Anand, Gagandeep S. and Breuval, Louise and Casertano, Stefano and Macri, Lucas M. and Li, Siyang and Yuan, Wenlong and Huang, Caroline D. and Jha, Saurabh and Murakami, Yukei S. and Beaton, Rachael and Brout, Dillon and Wu, Tianrui and Addison, Graeme E. and Bennett, Charles and Anderson, Richard I. and Filippenko, Alexei V. and Carr, Anthony},
title = {JWST Validates HST Distance Measurements: Selection of Supernova Subsample Explains Differences in JWST Estimates of Local H0},
journal = {The Astrophysical Journal}
}

@article{Freedman_2025,
doi = {10.3847/1538-4357/adce78},
url = {https://doi.org/10.3847/1538-4357/adce78},
year = {2025},
month = {may},
publisher = {The American Astronomical Society},
volume = {985},
number = {2},
pages = {203},
author = {Freedman, Wendy L. and Madore, Barry F. and Hoyt, Taylor J. and Jang, In Sung and Lee, Abigail J. and Owens, Kayla A.},
title = {Status Report on the Chicago-Carnegie Hubble Program (CCHP): Measurement of the Hubble Constant Using the Hubble and James Webb Space Telescopes},
journal = {The Astrophysical Journal}
}

@article{H0DN:2025lyy,
    author = "Casertano, Stefano and others",
    collaboration = "H0DN",
    title = "{The Local Distance Network: A community consensus report on the measurement of the Hubble constant at {\ensuremath{\sim}}1{\%} precision}",
    eprint = "2510.23823",
    archivePrefix = "arXiv",
    primaryClass = "astro-ph.CO",
    doi = "10.1051/0004-6361/202557993",
    journal = "Astron. Astrophys.",
    volume = "708",
    pages = "A166",
    year = "2026"
}

@article{Fields:2022mpw,
    author = "Fields, Brian D. and Olive, Keith A.",
    title = "{Implications of the non-observation of $^{6}$Li in halo stars for the primordial $^{7}$Li problem}",
    eprint = "2204.03167",
    archivePrefix = "arXiv",
    primaryClass = "astro-ph.GA",
    reportNumber = "UMN--TH--4118/22, FTPI--MINN--22/09",
    doi = "10.1088/1475-7516/2022/10/078",
    journal = "JCAP",
    volume = "10",
    pages = "078",
    year = "2022"
}

@article{Euclid:2025vml,
    author = "Frusciante, N. and others",
    collaboration = "Euclid",
    title = "{Euclid preparation. Review of forecast constraints on dark energy and modified gravity}",
    eprint = "2512.09748",
    archivePrefix = "arXiv",
    primaryClass = "astro-ph.CO",
    month = "12",
    year = "2025"
}

@article{scannapieco2012aquila,
  title={The Aquila comparison project: the effects of feedback and numerical methods on simulations of galaxy formation},
  author={Scannapieco, C et al and Wadepuhl, M and Parry, OH and Navarro, JF and Jenkins, Adrian and Springel, V and Teyssier, R and Carlson, E and Couchman, HMP and Crain, RA and others},
  journal={Monthly Notices of the Royal Astronomical Society},
  volume={423},
  number={2},
  pages={1726--1749},
  year={2012},
  publisher={Blackwell Publishing Ltd Oxford, UK}
}

@Article{galaxies5010017,
AUTHOR = {Del Popolo, Antonino and Le Delliou, Morgan},
TITLE = {Small Scale Problems of the $\Lambda$CDM Model: A Short Review},
JOURNAL = {Galaxies},
VOLUME = {5},
YEAR = {2017},
NUMBER = {1},
ARTICLE-NUMBER = {17},
URL = {https://www.mdpi.com/2075-4434/5/1/17},
ISSN = {2075-4434},
DOI = {10.3390/galaxies5010017}
}

@article{Ponomareva_2017,
   title={From light to baryonic mass: the effect of the stellar mass-to-light ratio on the Baryonic Tully–Fisher relation},
   volume={474},
   ISSN={1365-2966},
   url={http://dx.doi.org/10.1093/mnras/stx3066},
   DOI={10.1093/mnras/stx3066},
   number={4},
   journal={Monthly Notices of the Royal Astronomical Society},
   publisher={Oxford University Press (OUP)},
   author={Ponomareva, Anastasia A and Verheijen, Marc A W and Papastergis, Emmanouil and Bosma, Albert and Peletier, Reynier F},
   year={2017},
   month=Nov, pages={4366–4384} }

@article{DiCintio:2015eeq,
    author = "Di Cintio, Arianna and Lelli, Federico",
    title = "{The mass discrepancy acceleration relation in a $\Lambda$CDM context}",
    eprint = "1511.06616",
    archivePrefix = "arXiv",
    primaryClass = "astro-ph.GA",
    doi = "10.1093/mnrasl/slv185",
    journal = "Mon. Not. Roy. Astron. Soc.",
    volume = "456",
    number = "1",
    pages = "L127--L131",
    year = "2016"
}

@article{Lelli_2016,
doi = {10.3847/2041-8205/816/1/L14},
url = {https://doi.org/10.3847/2041-8205/816/1/L14},
year = {2016},
month = {jan},
publisher = {The American Astronomical Society},
volume = {816},
number = {1},
pages = {L14},
author = {Lelli, Federico and McGaugh, Stacy S. and Schombert, James M.},
title = {THE SMALL SCATTER OF THE BARYONIC TULLY–FISHER RELATION},
journal = {The Astrophysical Journal Letters}
}

@article{McGaugh_2012,
doi = {10.1088/0004-6256/143/2/40},
url = {https://doi.org/10.1088/0004-6256/143/2/40},
year = {2012},
month = {jan},
publisher = {The American Astronomical Society},
volume = {143},
number = {2},
pages = {40},
author = {McGaugh, Stacy S.},
title = {THE BARYONIC TULLY–FISHER RELATION OF GAS-RICH GALAXIES AS A TEST OF $\Lambda$CDM AND MOND},
journal = {The Astronomical Journal},
}

@article{Papastergis_2016,
   title={An accurate measurement of the baryonic Tully-Fisher relation with heavily gas-dominated ALFALFA galaxies},
   volume={593},
   ISSN={1432-0746},
   url={http://dx.doi.org/10.1051/0004-6361/201628410},
   DOI={10.1051/0004-6361/201628410},
   journal={Astronomy \& Astrophysics},
   publisher={EDP Sciences},
   author={Papastergis, E. and Adams, E. A. K. and van der Hulst, J. M.},
   year={2016},
   month=09, 
   pages={A39} }

@ARTICLE{Gambino2024,
       author = {{Gambino}, Serena and {Pace}, Francesco},
        title = "{Linear perturbations in Horndeski theories with spatial curvature}",
      journal = {arXiv},
         year = 2024,
        month = dec,
          eid = {arXiv:2412.01781},
        pages = {arXiv:2412.01781},
          doi = {10.48550/arXiv.2412.01781},
archivePrefix = {arXiv},
       eprint = {2412.01781},
 primaryClass = {gr-qc},
       adsurl = {https://ui.adsabs.harvard.edu/abs/2024arXiv241201781G}
}

@article{Tully:1977fu,
    author = "Tully, R. B. and Fisher, J. R.",
    title = "{A New method of determining distances to galaxies}",
    journal = "Astron. Astrophys.",
    volume = "54",
    pages = "661--673",
    year = "1977"
}

@misc{cirelli2025darkmatter,
      title={Dark Matter}, 
      author={Marco Cirelli and Alessandro Strumia and Jure Zupan},
      year={2025},
      eprint={2406.01705},
      archivePrefix={arXiv},
      primaryClass={hep-ph},
      url={https://arxiv.org/abs/2406.01705}, 
}

@article{Misiaszek_2024,
   title={Direct Detection of Dark Matter: A Critical Review},
   volume={16},
   ISSN={2073-8994},
   url={http://dx.doi.org/10.3390/sym16020201},
   DOI={10.3390/sym16020201},
   number={2},
   journal={Symmetry},
   publisher={MDPI AG},
   author={Misiaszek, Marcin and Rossi, Nicola},
   year={2024},
   month=Feb, pages={201} }

@article{Cebri_n_2023,
   title={Review on dark matter searches},
   volume={2502},
   ISSN={1742-6596},
   url={http://dx.doi.org/10.1088/1742-6596/2502/1/012004},
   DOI={10.1088/1742-6596/2502/1/012004},
   number={1},
   journal={Journal of Physics: Conference Series},
   publisher={IOP Publishing},
   author={Cebrián, Susana},
   year={2023},
   month=May, pages={012004} }

@article{Gupta:2023,
author = {Gupta, Rajendra},
year = {2023},
month = {07},
pages = {3385-3395},
title = {JWST early Universe observations and $\Lambda$CDM cosmology},
volume = {524},
journal = {Monthly Notices of the Royal Astronomical Society},
doi = {10.1093/mnras/stad2032}
}

@ARTICLE{Lelli2026,
    author = {Lelli, Federico and  al. },
    title="{The MOND research program: A white paper on status and prospects}", 
    journal = {{\rm to be published}},
    year = 2026
}

\end{document}